%% file: main.tex
\documentclass[acmtog, nonacm]{acmart}
\AtBeginDocument{%
  }

\setcopyright{acmlicensed}
\copyrightyear{2018}
\acmYear{2018}
\acmDOI{XXXXXXX.XXXXXXX}
\acmConference[Conference acronym 'XX]{Make sure to enter the correct
  conference title from your rights confirmation email}{June 03--05,
  2018}{Woodstock, NY}
\acmISBN{978-1-4503-XXXX-X/2018/06}

\usepackage[boxed,ruled,linesnumbered]{algorithm2e} %
\let\oldnl\nl%
\newcommand{\nonl}{\renewcommand{\nl}{\let\nl\oldnl}}%

\SetAlFnt{\small}
\SetAlCapFnt{\small}
\SetAlCapNameFnt{\small}
\SetAlCapHSkip{0pt}
\IncMargin{-\parindent}
\newlength\savedwidth

\usepackage{enumitem} 
\usepackage{color, colortbl}

\DeclareMathOperator*{\argmin}{\arg\!\min}
\usepackage{soul}
\usepackage[normalem]{ulem} %
\usepackage{multirow}
\usepackage{bbding}
\usepackage{amsmath}
\usepackage{subcaption}
\usepackage{xcolor,colortbl}
\definecolor{headercolor}{HTML}{b2df8a}
\definecolor{header2base}{HTML}{AE2573}
\colorlet{header2color}{header2base!30}  %
\usepackage[dvipsnames]{xcolor}
\input{newcommands}
\usepackage{wrapfig}

\begin{document}

\title{VoroUDF: Meshing Unsigned Distance Fields with Voronoi Optimization}

\author{Ningna Wang}
\affiliation{%
  \institution{Columbia University}
  \state{New York}
  \country{USA}
}
\email{ningna.wang@columbia.edu}

\author{Zilong Wang}
\affiliation{%
  \institution{University of Texas at Dallas}
  \city{Richardson}
  \country{USA}
}
\email{Zilong.Wang@utdallas.edu}

\author{Xiana Carrera}
\affiliation{%
  \institution{Columbia University}
  \city{New York}
  \country{USA}
}
\email{x.carrera@columbia.edu}

\author{Xiaohu Guo}
\affiliation{%
  \institution{University of Texas at Dallas}
  \city{Richardson}
  \country{USA}
}
\email{xguo@utdallas.edu}

\author{Silvia Sellán}
\affiliation{%
  \institution{Columbia University}
  \city{New York}
  \country{USA}
}
\email{silviasellan@cs.columbia.edu}

\renewcommand\shortauthors{Wang et. al}

\begin{abstract}
We present VoroUDF, an algorithm for reconstructing high-quality triangle meshes from Unsigned Distance Fields (UDFs). Our algorithm supports non-manifold geometry, sharp features, and open boundaries, without relying on error-prone inside/outside estimation, restrictive look-up tables nor topologically noisy optimization. \revision{Unlike fixed-grid approaches, our Voronoi-based formulation optimizes a set of movable seeds that travel freely along the iso-surface, jointly minimizing a tangent-plane fitting energy, driven by a fixed set of surface samples queried once from the UDF and its gradient, and a repulsion energy that keeps the seeds evenly distributed.} It achieves significantly improved topological consistency and geometric fidelity compared to existing methods, while producing lightweight meshes suitable for downstream real-time and interactive applications.
\end{abstract}

\begin{teaserfigure}
  \includegraphics[width=\textwidth]{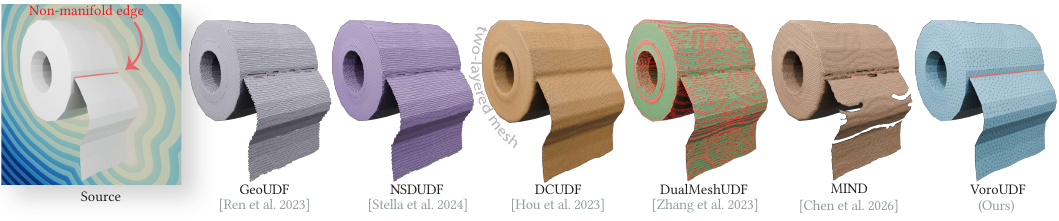}
  \vspace{-0.7cm}
  \caption{\rewritten{We propose an algorithm for reconstructing surfaces with non-manifold structures (highlighted in red), sharp features, and open boundaries from unsigned distance fields with gradient information.}}
  \label{fig:teaser}
\end{teaserfigure}

\maketitle

\input{secs/1_intro}

\input{secs/2_related}

\input{secs/3_1_method}
\input{secs/4_results}

\input{secs/5_comparisons}

\input{secs/6_application}
\input{secs/7_conclusion}

\begin{acks}
We sincerely thank all anonymous reviewers for their constructive feedback. We also want to thank Jiayi (Eris) Zhang for sharing their models used in \autoref{fig:drape_sculpt} and \autoref{fig:modeling}.
The Geometry and the City lab at Columbia University is supported by generous gifts from nTop, Adobe, Dandy and Braid Technologies, as well as by research sponsorships from Meshy, DreamSports, and a grant from the Columbia Engineering Interdisciplinary Research Fund.
\end{acks}

\bibliographystyle{ACM-Reference-Format}
\bibliography{reference}

\appendix

\clearpage
\input{secs/10_app.tex}

\end{document}

%% file: newcommands.tex
\theoremstyle{definition}

\newcommand{\isosurface}{\chi}
\newcommand{\mesh}{\mathcal{M}}
\newcommand{\allseeds}{\mathcal{X}}
\newcommand{\alltris}{\mathcal{F}}
\newcommand{\udf}{d}
\newcommand{\udfunc}{F}
\newcommand{\udfgrad}{\mathbf{n}}
\newcommand{\sample}{\mathbf{p}}
\newcommand{\seed}{\mathbf{x}}

\newcommand{\dtan}{D_{tan}}
\newcommand{\dcov}{D_{cov}}
\newcommand{\thredudf}{\epsilon_{udf}}
\newcommand{\thredudfgrad}{\epsilon_{grad}}
\newcommand{\kernel}{\sigma}
\newcommand{\energy}{E}

\newcommand{\meshrecon}{\mathcal{M}_{r}}

\newcommand{\dimension}{\mathbb{R}}
\newcommand{\td}{\textit{TE}}
\newcommand{\hd}{\textit{HE}}
\newcommand{\cd}{\textit{CE}}
\newcommand{\ecd}{\textit{ECE}}
\newcommand{\nmcd}{\textit{NM-CE}}
\newcommand{\triangleQ}{\textit{TQ}}
\newcommand{\best}[1]{\underline{\textbf{#1}}}
\newcommand{\second}[1]{\textbf{#1}}

\newcommand{\ie}{\textit{i.e., }}
\newcommand{\eg}{\textit{e.g., }}

\newcommand{\revision}[1]{{\color{black}#1}}

\newcommand{\revisionenv}{\color{black}}
\newcommand{\rewritten}[1]{{\color{black}#1}}

\usepackage{layouts}
\usepackage{calc}
\makeatletter
\newcommand{\layoutdetails}{%
\begin{tabular}{ll}
\texttt{\textbackslash{textwidth}} & \printinunitsof{in}\prntlen{\textwidth} \\
\texttt{\textbackslash{linewidth}} & \printinunitsof{in}\prntlen{\linewidth} \\
Main text font &  \f@size pt \f@family \\
\sffamily \small Caption text font &  \sffamily \small \f@size pt \f@family \\
\end{tabular}%
}
\makeatother

%% file: secs/1_intro.tex
\section{Introduction}

\begin{figure*}[t!]
    \centering
    \includegraphics[width=\linewidth]{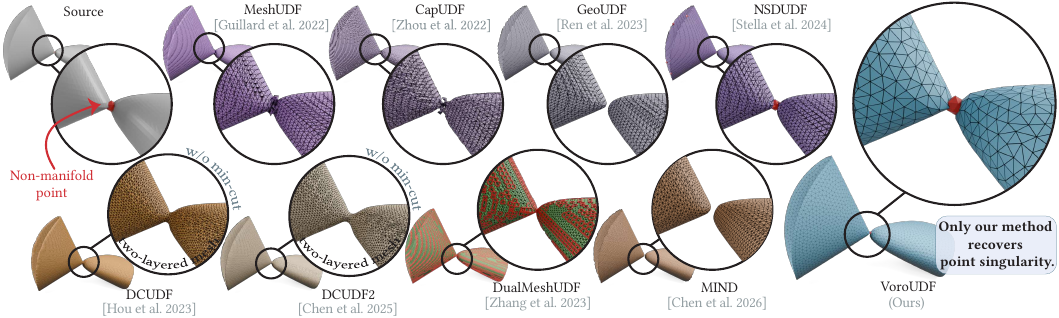}
    \vspace{-0.7cm}
    \caption{Unlike marching-cube-based methods~\cite{guillard2022meshudf, zhou2022capudf, ren2023geoudf, stella2024nsdudf} that rely on pseudo-sign estimation, double-covering-based methods~\cite{hou2023dcudf, chen2025dcudf2} that produce double-layered meshes, and dual-contouring-based methods~\cite{zhang2023dualmeshUDF} that introduce erroneous connectivity, \revision{material-interface-based method~MIND~\cite{chen2026mind} that outputs disjunct interfaces,} our method is the only approach capable of reconstructing non-manifold structures (highlighted in red). }
    \label{fig:rw_nonmanifold}
\end{figure*}

\textit{Unsigned Distance Fields} (UDFs) have emerged as a powerful representation for shapes of arbitrary topology containing sharp and even non-manifold features. 
Their ability to represent thin, geometrically challenging self-intersecting structures has made them popular in artistic 3D modeling applications (see \autoref{fig:teaser}), while their flexibility has increased their adoption in learning-based shape representations and reconstruction pipelines for real-world geometry.
\revision{Reconstructing lightweight high-quality surface meshes with non-manifold structures, open boundaries, and sharp features from UDFs is critical for efficient downstream applications such as texture mapping, physical simulation, animation, and geometric processing.}

However, converting a UDF into an explicit surface mesh remains challenging. Without a consistent sign to differentiate inside from outside, classic grid-based iso-surfacing techniques \cite{marchingcube,ju2002dual} cannot be directly applied.
Methods that combine sign estimation with Marching Cubes lookup tables \cite{guillard2022meshudf,zhou2022capudf,ren2023geoudf,stella2024nsdudf}
are highly unstable around thin regions and fail to reconstruct non-manifold structures (see \autoref{fig:teaser}, second and third from left).
Dual grid contouring approaches with purpose-made quadratic optimizations \cite{zhang2023dualmeshUDF} produce topological noise and undesirable non-manifold edges missing from the input (\autoref{fig:teaser}, \revision{third from right}).
When non-manifold features are intentionally present in the input (\eg \autoref{fig:teaser}, left), algorithms based on contracting a surface mesh of a volumetric offset \cite{hou2023dcudf, chen2025dcudf2} cannot recover them by design (\autoref{fig:teaser}, \revision{fourth from right}).

\revision{In this paper, we introduce VoroUDF, a Voronoi-based optimization framework for extracting iso-surfaces with non-manifold structures, open boundaries, and sharp features from UDFs. 
Our key insight is that the failure of prior methods to reconstruct fine features is due to their reliance on fixed space discretizations; i.e., their \emph{Eulerian} nature.
By contrast, we propose a 
\emph{Lagrangian} method in which a set of movable \emph{seeds} are aligned with the shape's sharp edges, open boundaries, and non-manifold junctions.
The position of these seeds is optimized iteratively to balance feature alignment, coverage and triangle quality, as the seeds serve as vertices of the final extracted mesh.}

\revision{Our flagship result is a comprehensive qualitative comparison to prior work on three commonly-used benchmarks containing sharp edges~\cite{koch2019abc}, non-manifold features~\cite{ren2025sals} and open boundaries~\cite{zhu2020deepfashion}, as well as on one purpose-made dataset containing synthetic non-manifold features.
Throughout all tests, at a low to medium output mesh resolution, our method significantly outperforms every prior work on both general accuracy and feature alignment (see \autoref{fig:teaser}).
Indeed, to match our method's accuracy, previous methods require significantly higher resolutions, making their outputs less suited for the real-time and interactive downstream applications that we prototype throughout the paper (\eg Figures \ref{fig:drape_sculpt}, \ref{fig:modeling}, \ref{fig:table}, \ref{fig:swept_envelope})}

\revision{Beyond showcasing our method's performance on its intended task, we also explore its usefulness in a broader set of applications. For example, while our method assumes the input to be a continuous, exact UDF, we also measure its performance on both neural and noisy inputs. 
Finally, we consider point clouds as inputs and use our method to reconstruct single-layered, lightweight and open meshes from incomplete or interior scans.}

%% file: secs/2_related.tex
\section{Related Work}

UDFs can represent surfaces of arbitrary topology, making them particularly valuable in domains such as artistic modeling, manufacturing, and industrial design, where accurate recovery of complex surface structures is critical. Owing to this flexibility, UDFs have become a powerful tool across a wide range of scientific fields—from medical imaging~\cite{jin2024misner, sorensen2022nudf} to geology~\cite{zhang2026mb} and additive manufacturing~\cite{brunton2021displaced}. In computer graphics, their expressive capacity has been extensively utilized in geometric deep learning and generative 3D modeling~\cite{xu2025details, chibane2020NDF, liu2023neudf, long2023neuraludf, zhao2021learning}.

Despite their expressiveness, UDFs remain poorly suited for many downstream applications due to their implicit nature. Tasks such as finite element simulation, texture mapping, real-time rendering, and animation often require converting UDFs into explicit surface representations like polygonal meshes. This conversion is nontrivial: UDFs lack orientation and sign information, making direct mesh extraction challenging. \revision{Two directions have emerged in response. One sidesteps the meshing problem by changing the representation, \eg SALS~\cite{ren2025sals} attains non-manifold reconstructions by fitting a learned line-segment representation to the input. The other takes the UDF as given and aims for a high-quality mesh of its zero level set; our method belongs to this second direction, and we also adopt the SALS dataset of intersecting shapes to evaluate non-manifold reconstruction (Section~\ref{sec:quatitative_results}). In the following, we review the main strategies for robust UDF-to-mesh reconstruction, each with important limitations as well as their strengths and weaknesses.}

\subsection{Reconstruction from UDFs}

\begin{figure}
    \centering
    \begin{subfigure}[b]{\linewidth}
        \centering
        \includegraphics[width=\linewidth]{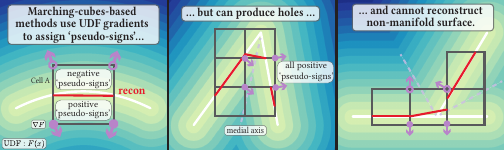}
        \vspace{-8pt}
    \end{subfigure}
    \noindent\rule[10pt]{\linewidth}{1pt} 
    \begin{subfigure}[b]{\linewidth}
        \centering
        \vspace{-8pt}
        \includegraphics[width=\linewidth]{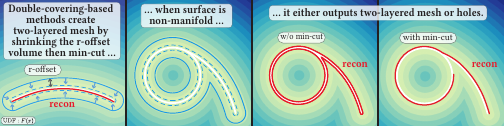}
        \vspace{-8pt}
    \end{subfigure}
    \noindent\rule[10pt]{\linewidth}{1pt} 
    \begin{subfigure}[b]{\linewidth}
        \centering
        \vspace{-8pt}
        \includegraphics[width=\linewidth]{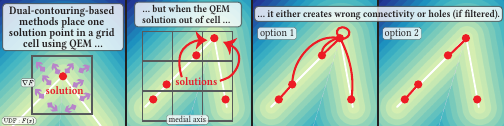}
    \end{subfigure}
    \vspace{-0.7cm}
    \caption{Illustration of why marching-cubes-based methods (top row), double-covering-based methods (middle row), and dual-contouring-based methods (bottom row) fail to preserve the non-manifoldness of the iso-surface.}
    \label{fig:why_not}
\end{figure}

\paragraph{Marching-cubes-based methods}
One family of methods adapts grid-based isosurface extraction (\eg Marching Cubes variants) to UDFs by introducing a pseudo-sign per grid cell~\cite{guillard2022meshudf, zhou2022capudf, ren2023geoudf}. These approaches attempt to infer on which side of the surface each voxel corner lies (since all UDF values are positive) by analyzing the UDF’s gradients or other heuristics. Essentially, they locally treat the UDF as a signed field and then apply the Marching Cubes lookup to extract a mesh. This works only under a key assumption: each grid cell that contains the surface must lie entirely on one side of the surface’s medial axis. If a cell spans across the medial axis, the pseudo-sign becomes ill-defined and unstable (see Figure~\ref{fig:why_not} top row). 
The learning-based method NSDUDF~\cite{stella2024nsdudf} predicts pseudo-sign information from local views, however, it struggles to generalize and may fail to produce accurate results on previously unseen shapes or topologies, particularly in regions with non-manifold geometry.

\paragraph{Double-covering-based methods}
Another approach is to temporarily convert the open or non-manifold surface into a closed manifold by using a double covering technique~\cite{hou2023dcudf, chen2025dcudf2}. This approach extracts a thin offset surface around the target shape, which forms a closed, orientable manifold, and then optimizes this double-layered surface onto the direction of the isosurface. This yields a valid mesh even when the input surface is open. If the input surface is truly non-manifold, however, the method cannot produce a single unified mesh, as it instead retains two overlapping surface layers at the non-manifold regions (see Figure~\ref{fig:why_not} middle row).

\paragraph{Dual-contouring-based methods}
To avoid relying on a global pseudo-sign, researchers have also proposed dual-contouring-based methods for UDF meshing.
The learning-based method NDC~\cite{chen2022neuraldc} enhances the classic Dual Contouring algorithm by refining UDFs using learned priors; however, it struggles to generalize to shapes with arbitrary topology and unseen geometric configurations.
A recent example is DualMeshUDF~\cite{zhang2023dualmeshUDF}, which uses an adaptive octree and computes each mesh vertex by solving a quadratic error function (QEF) that fits the local unsigned field (analogous to finding the intersection of estimated tangent planes). These approaches can directly place vertices on the surface without requiring sign flips, enabling it to capture sharp edges and even non-manifold structures that Marching Cubes variants might miss. However, ensuring topologically correct connectivity becomes challenging for such dual methods, as they all rely on the assumption that the solution space will not be far from the grid cell and then minimizes towards a point inside of the cell close to the solution. The erroneous connectivity appears when the reconstructed surface point computed for an octree cell lies outside the cell. If these noisy points are discarded or clamped, the mesh quality is  compromised by holes (see \autoref{fig:why_not} bottom row). In practice, DualMeshUDF filters out vertices that fall outside their cell, which means those cells end up with no mesh vertex and thereby produce a hole in the reconstructed mesh. This situation tends to occur when a non-empty cell covers surface regions on both sides of its medial axis, causing the QEF solution to drift out of bounds. One might refine the octree further to alleviate this issue, but increasing resolution explosively raises the mesh complexity. For example, in our experiments the Klein bottle model (see \autoref{fig:method_thinning_vs_dmudf}) required over $13k$ vertices with DualMeshUDF to avoid holes, whereas a far coarser mesh (\revision{roughly $5.2k$ vertices}) suffices with our method. Moreover, DualMeshUDF and similar octree methods impose a maximum depth limit, so they cannot indefinitely subdivide to resolve every ambiguous cell, occasionally leading to small holes or missed facets.

\begin{figure}
    \centering
    \includegraphics[width=\linewidth]{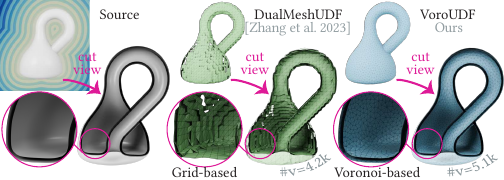}
    \vspace{-15pt}
    \caption{Dual-contouring-based method~\cite{zhang2023dualmeshUDF} uses voxel grids struggles to reconstruct UDF iso-surfaces with thin layers, often resulting in zig-zag connectivity and holes when using a relatively small number of vertices. In contrast, our method, based on a Voronoi partition, avoids these issues and produces clean, consistent connectivity.}
    \label{fig:method_vs_dmudf}
\end{figure}

\begin{figure}
    \centering
    \includegraphics[width=\linewidth]{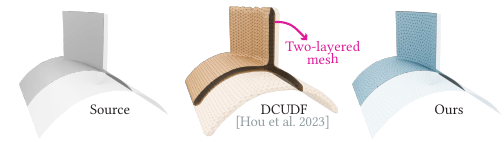}
    \vspace{-20pt}
    \caption{\rewritten{Our method successfully reconstructs non-manifold structures from UDFs, whereas the double-covering-based method~\cite{hou2023dcudf} returns a two-layered mesh around non-manifold regions.}}
    \label{fig:method_vs_dcudf}
\end{figure}

\begin{figure}
    \centering
    \includegraphics[width=\linewidth]{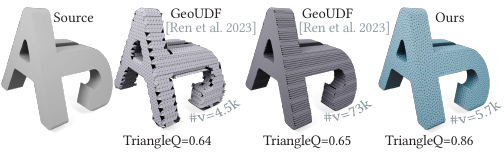}
    \vspace{-20pt}
    \caption{\rewritten{Our method reconstructs the UDF iso-surface with better triangle quality compared to the marching-cubes-based approach~\cite{ren2023geoudf}.}}
    \label{fig:method_vs_mc}
\end{figure}

\begin{figure*}[t!]
    \centering
    \IfFileExists{figs/pipeline.pdf}{%
        \includegraphics[width=\linewidth]{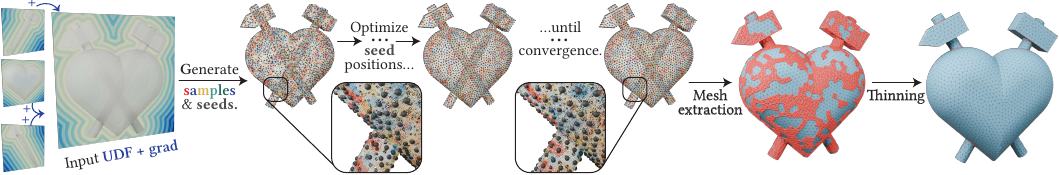}%
    }{%
        \fbox{\parbox[c][1.2in][c]{0.95\linewidth}{\centering
        \texttt{figs/pipeline.pdf} placeholder (F1)}}%
    }
    \vspace{-0.7cm}
    \caption{\revision{Instead of relying on a fixed discretization, our method uses a set of \emph{seeds} that are initialized on the surface (\autoref{sec:method}) and their positions optimized to balance distribution and feature alignment (\autoref{sec:energy}). The seeds become the vertices of the output mesh (\autoref{sec:mesh_extraction}), which is cleaned up to produced our final output.}}
    \label{fig:pipeline}
\end{figure*}

\revision{
The grid-based methods above are all \emph{Eulerian}: their vertices are confined to the cells of a fixed grid, so the resolution of the output mesh is inherited from that of the discretization, and capturing fine features means refining the grid. Our method is instead \emph{Lagrangian}: its vertices are movable \emph{seeds} that travel along the iso-surface itself, settle exactly on sharp edges, open boundaries, and non-manifold junctions, and their budget alone sets the output resolution. This decoupling of the output density from the input discretization takes inspiration from Reach for the Spheres~\cite{sellan2023rfs} and Reach for the Arcs~\cite{sellan2024rfa}, which demonstrated its benefits for \emph{signed} fields and watertight manifold shapes.
}

\revision{
\paragraph{Concurrent work}
For completeness, we also position our method against three concurrent publications. MIND~\cite{chen2026mind} extracts non-manifold meshes from UDFs by generating material interfaces from a volumetric multi-label assignment; as the approach closest to ours in goal, we compare against it in \autoref{fig:teaser}, \ref{fig:rw_nonmanifold} and Section~\ref{sec:comp}. 
DCx~\cite{bao2026dual} extends dual contouring with expanded cubes to extract zero level sets of neural UDFs, and OffsetAxis~\cite{huang2026offsetaxis} meshes a UDF by extracting the medial axis of a thin offset volume, which, like our output, can carry non-manifold structure. Neither implementation is publicly available at the time of writing.
}

\revision{
\subsection{Point Representations}
Our method internally operates on point sets: a dense, fixed set of surface \emph{samples} guides a coarse set of movable \emph{seeds} (Section~\ref{sec:sampling}). For completeness, we discuss the two directions of prior work this design touches: methods that, like our seed optimization, place and optimize points on an implicit surface, and methods that reconstruct a mesh from point samples, against which we compare in Section~\ref{sec:comp_to_point_cloud}.

\paragraph{Point-based sampling of implicit surfaces}
A long line of work places and optimizes particles on implicit surfaces: physically-based polygonization~\cite{figueiredo1992physically}, repulsion-based particle systems~\cite{witkin1994using, levet2006fast}, sampling of point-set implicits~\cite{proenca2007sampling}, and, more recently, hybrid point/implicit representations for neural fields~\cite{atzmon2019controlling, yifan2021isopoints}. Related point-resampling techniques such as LOP~\cite{lipman2007lop}, its continuous variant CLOP~\cite{preiner2014clop}, and edge-aware resampling~\cite{huang2013edgeaware} redistribute a point cloud to be uniform or feature-aware. Our seed optimization is related in spirit: like these methods, we advect a set of points under attraction and distribution forces. Our goal, however, is not the sampling itself but rather let the advected seeds become the feature-aligned vertices of the final reconstructed mesh.

\paragraph{Surface reconstruction from point samples}
Since our method internally builds a dense point sampling of the iso-surface, one may ask whether an off-the-shelf point-cloud reconstruction could replace our optimization and extraction. Classic Delaunay-based methods offer provable sampling and meshing guarantees~\cite{boissonnat2005provably}, restricted-Delaunay reconstruction has been extended with manifoldness-repair strategies~\cite{wang2022rdt}, and Screened Poisson reconstruction~\cite{kazhdan2013screened} as well as orientation-recovery methods such as Signing the Unsigned~\cite{mullen2010signing} estimate or require a consistent global orientation, an assumption which does not hold for the open and non-manifold surfaces we target. We examine this connection in both directions. First, we feed the very same samples our pipeline generates to three classical point-cloud reconstruction methods and compare their outputs against ours (\autoref{fig:comp_pointcloud}, Section~\ref{sec:comp_to_point_cloud}). Second, a raw point cloud can also serve as our input, with distances and gradients queried directly from the cloud; we demonstrate this on three scanned scenes from \citet{zhou2013dense} (\autoref{fig:comp_pointcloud_inputs}, Section~\ref{sec:comp_to_point_cloud}).
}

%% file: secs/3_1_method.tex
\section{Method}
\label{sec:method}
\label{sec:sampling}

The input to our method is a continuous UDF that provides both unsigned distance values $\udf = \udfunc(\seed)$ given a point $\seed \in \dimension^3$, which measures proximity to the unknown iso-surface $\isosurface$, and gradients $\udfgrad(\seed) = \nabla \udfunc(\seed) / \|\nabla \udfunc(\seed)\|$, which indicate the direction of value increase. In this section, we introduce an iterative optimization algorithm to compute a high-quality triangle mesh $\mesh = (\allseeds, \alltris)$ that approximates $\isosurface$ while faithfully capturing non-manifold structures, open boundaries, and sharp features.
\revision{Our method operates on two point sets with distinct roles: \emph{seeds} and \emph{samples}.}

\paragraph{\revision{Seeds}}
\revision{A coarse set of $N$ movable \emph{seeds} $\{\seed_i\}_{i=1}^N$, whose positions are the variables of our optimization: they travel freely along the iso-surface and become the vertices of the output mesh. We initialize them directly on $\isosurface$ by casting random rays through the bounding volume and locating their intersections with the zero level set of $\udfunc$ by sphere tracing along each ray~\cite{ling2025uniform}; see \autoref{fig:pipeline}~(b), black points.
}

\paragraph{\revision{Samples}}
\revision{
A much denser, \emph{fixed} set of $M$ surface \emph{samples} $\sample_m$ (see \autoref{fig:pipeline}~(b), colored points), generated with
two complementary strategies.
First, we combine ray casting \cite{ling2025uniform} with Poisson disk decimation, using an ideal separation $r$ (see Section~\ref{sec:parameters} for details).
Second, we draw uniform random points in the bounding volume of the shape, and project each onto its closest surface point.
Both type of samples are also equipped with a surface normal $\udfgrad_m$, given by querying the last sample during sphere tracing or the pre-projection sample, respectively.
The proportion between sample types is ablated in \autoref{fig:ablation_ratio_of_grid_samples} (in practice, we use $70\%$ Poisson samples).
}

\revision{
Our method will exploit these two types of point sets to produce its reconstruction. First, the \emph{samples} will be assigned to corresponding \emph{seeds} and used to construct and solve an optimization problem (see Section~\ref{sec:energy}), the output of which will be a new set of \emph{seed} positions.
Then, these same seeds will become the vertices of the output mesh, which is extracted and repaired using an approximate geodesic Voronoi partition (see Section~\ref{sec:mesh_extraction}).
Our full algorithm is summarized in \autoref{fig:pipeline}, and detailed in Algorithm~\ref{alg:opt},\ref{alg:extract}.
}

\begin{figure}[t]
    \centering
    \includegraphics[width=\linewidth]{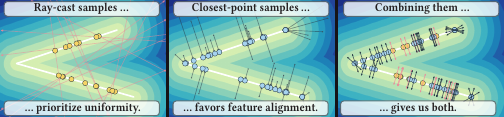}
    \vspace{-0.7cm}
    \caption{\revision{We use a combination of uniform and closest point samples to ensure coverage and feature alignment.}}
    \label{fig:method_two_samples}
\end{figure}

\subsection{\revision{Joint Repulsive--Tangent Energy}}
\label{sec:energy}

\revision{We now build an energy whose optimization variables are the seed positions $\allseeds = \{\seed_i\}_{i=1}^N$. The seeds must satisfy two conditions. First, since the seeds become the vertices of the output mesh, they should be \emph{evenly distributed} over the surface. Second, they should \emph{align with the features} of $\isosurface$: sharp edges, open boundaries, and non-manifold junctions. Our energy therefore has two terms, one for each goal.}

\revision{To keep the seeds evenly distributed, which outputs meshes with higher triangle quality, we adopt a standard Gaussian particle model \cite{witkin1994using, zhong2013particle, wang2025matstruct} to build a \emph{repulsive} energy
\begin{equation}
\energy_{rep}(\allseeds) = \sum_{i=1}^{N} \sum_{j \in \mathcal{N}(i)} \exp\!\left(-\frac{\lVert \seed_i - \seed_j \rVert^2}{4\kernel^2}\right), \label{eqn:energy_rep}
\end{equation}
where the optimal kernel width $\kernel$ is estimated once for each shape (see Section~\ref{sec:parameters}), and $\mathcal{N}(i)$ are the seeds falling within $5\kernel$ of $\seed_i$.
}

\revision{
\begin{wrapfigure}[7]{r}{1in}
    \hspace{-10pt} %
    \includegraphics[width=1in]{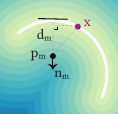}
\end{wrapfigure}  
To encourage feature alignment, we will make use of the samples $\sample_m$. Each of them is assigned to a single seed, denoted $a(m) \in \{1, \dots, N\}$ (we discuss strategies for this assignment in Section~\ref{sec:partitions}).
Each sample carries an estimate of the local geometry of $\isosurface$: a sample $\sample_m$ with unit gradient $\udfgrad_m$ and distance $\udf_m = \udfunc(\sample_m)$ defines an estimated tangent plane, and any surface point $\seed$ near $\sample_m$ satisfies $\udfgrad_{m}^{\top} (\sample_{m} - \seed) - \udf_{m} \approx 0$ (see inset). Each sample thus measures a \emph{tangent-plane distance} to its assigned seed $\seed$, the squared deviation from its estimated tangent plane:
\begin{equation}
\dtan(\seed, \sample_{m}, \udfgrad_{m}) = \Big( \udfgrad_{m}^{\top} (\sample_{m} - \seed) - \udf_{m} \Big)^2, \label{eqn:dist_tan}
\end{equation}
leading us to define the \emph{tangent} energy term
\begin{equation}
\energy_{tan}(\allseeds) = \sum_{m=1}^{M} \dtan\big(\seed_{a(m)}, \sample_{m}, \udfgrad_{m}\big). \label{eqn:energy_tan}
\end{equation}
This term pulls each seed toward the surface inferred from its assigned samples.
Even better, in the neighborhood of a sharp feature, boundary or non-manifold component, it will encourage the seed to concentrate on the feature itself.
This property of this tangent energy has been exploited before; for example, by dual contouring~\cite{ju2002dual, schaefer2002dual} in the context of signed distance data and by DualMeshUDF~\cite{zhang2023dualmeshUDF} for UDF inputs. There, however, each vertex minimizes the energy of the samples inside a \emph{fixed} grid cell: the minimizer frequently drifts outside its cell (\eg whenever the cell spans the medial axis, see Figure~\ref{fig:why_not}), and discarding or clamping such vertices produces holes and erroneous connectivity. Our seeds are bound to no cell and simply move freely where their assigned samples send them.}

\revision{Put together, we combine \autoref{eqn:energy_rep} and \autoref{eqn:energy_tan}:
\begin{equation}
\energy(\allseeds) = \energy_{tan}(\allseeds) + \omega\, \energy_{rep}(\allseeds),
\label{eqn:energy_total}
\end{equation}
where the choice of the $\omega$ is discussed in Section \ref{sec:parameters} and the full energy is minimized using L-BFGS (each seed $\seed$ is re-projected to the zero level-set using its gradient after each L-BFGS iteration).
The optimization terminates when either a relative tolerance ($10^{-8}$) or a maximum iteration number ($200$) are reached.
}

\begin{figure}[t]
    \centering
    \includegraphics[width=\linewidth]{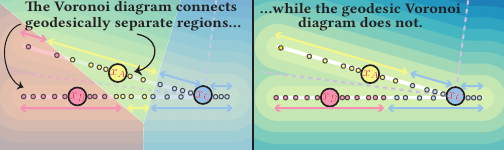}
    \vspace{-15pt}
    \caption{\rewritten{A Voronoi diagram with three mutually adjacent seeds. $\mathbf{x_A}$ and $\mathbf{x_B}$ are not adjacent in the \emph{geodesic} Voronoi diagram, so the dual edge between them does not exist in the final reconstructed mesh.}}
    \label{fig:method_gvd}
\end{figure}

\paragraph{Assigning samples to seeds.}
\label{sec:partitions}
\revision{
Each sample is assigned to one seed, and the assignment is recomputed at every L-BFGS iteration. A natural choice is to assign each sample to its closest seed in Euclidean distance by minimizing $\dcov(\seed, \sample_{m}) = \big\lVert \seed - \sample_{m} \big\rVert_2^2$
which yields exactly the Voronoi diagram of the seeds. However, this lets spatially close but geodesically distant regions affect each other (see \autoref{fig:method_gvd}). The ideal remedy would be the \emph{geodesic} Voronoi diagram (GVD). We do use an
approximation of it for mesh reconstruction (Section \ref{sec:mesh_extraction}), but its cost is too high to recompute at every iteration during optimization. We therefore settle on a middle ground: we augment the Euclidean distance with the tangent-term (\autoref{eqn:dist_tan}), and assign each sample to the seed minimizing the combined \emph{tangency-aware Voronoi distance}:
\begin{equation}
a(m) = \argmin_{i \,:\, \seed_i \in \mathrm{kNN}(\sample_m)} \; \dtan(\seed_i, \sample_{m}, \udfgrad_{m}) + \eta\, \dcov(\seed_i, \sample_{m}),
\label{eqn:tangency_aware_voronoi}
\end{equation}
where $\eta = 1$ in all experiments.
}

\begin{figure}[t]
    \centering
    \includegraphics[width=\linewidth]{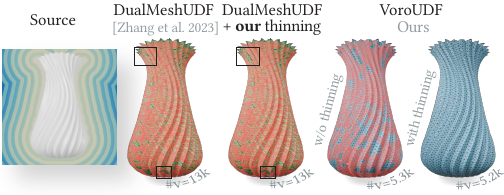}
    \vspace{-20pt}
    \caption{\rewritten{We apply our thinning method described in Section~\ref{sec:thinning} to both our output and that of DualMeshUDF~\cite{zhang2023dualmeshUDF}, which contains no tetrahedral cells and only a few polyhedral cells that are identified and removed. Non-manifold regions are highlighted in red.}}
    \label{fig:method_thinning_vs_dmudf}
\end{figure}

\subsection{\rewritten{Mesh Extraction}}
\label{sec:mesh_extraction}

After optimizing the seeds, the next step is to connect them to form a triangle mesh.
\rewritten{As discussed above, we use a mesh extraction} method based on the \textit{Geodesic Voronoi Diagram} (GVD) (see \autoref{fig:method_gvd}).

Computing an exact GVD on an implicit zero-level set (without an explicit mesh) is challenging, so we develop a sampling-based approximation, \rewritten{for which we re-use the Poisson disk samples $\sample_m$ from \autoref{sec:method}.}
We first construct a $k$-nearest neighbor (KNN) graph on these samples. Next, we prune edges from this graph that likely do not lie on the surface. In particular, we remove any edge if either (1) the unsigned distance value at its midpoint is above a threshold (\ie $\udfunc \big(1/2(\seed_i + \seed_j)\big) > \thredudf = 0.1$), indicating the midpoint is off the surface, or (2) the UDF gradient at either endpoint is too aligned with the edge’s direction (quantified by the dot product exceeding $\thredudfgrad=0.3$), indicating the edge is not tangent to the surface. After this filtering, we treat each seed as a source and run a \rewritten{Dijkstra sweeping algorithm, inspired by \cite{xin2022surfacevoronoi}, on the proximity graph} to label every sample with its nearest seed in terms of geodesic distance, effectively simulating a flooding process on the surface. Finally, we construct the triangle connectivity (the dual of the approximate GVD) by examining the labeled samples: for each surface sample labeled $i$ (nearest to seed $\seed_i$), if it has neighboring samples labeled $j$ and $k$ (with $j \neq i$ and $k \neq i$), we introduce a triangle connecting seeds $(\seed_i, \seed_j, \seed_k)$ in the mesh. Each such triangle corresponds to a junction where three Voronoi regions meet on the surface.

\input{tables/tab_ablation_cvt_vs_repulsive_fixed_vs_decay_SALS_reconGVD.tex}

\paragraph{Thinning}
\label{sec:thinning}

The initial mesh produced by the geodesic Voronoi diagram (GVD) reconstruction is inherently \textit{thick}, as the dual of the GVD introduces flat but solid volumetric cells. As a result, the extracted surface is not yet a thin manifold. 
\revision{Since our method is designed to reconstruct surfaces containing \emph{purposeful} non-manifold and volumetric components, any strategy to remove these must rely on prior assumptions about the input shape. 
We thus combine heuristics based on the size and shape of the desired and undesired features into} a \textit{thinning} procedure inspired by~\citet{2022MATFP, wang2024mattopo, wang2025matstruct}.

\begin{wrapfigure}[7]{r}{1.1in}
    \vspace{-20pt} %
    \hspace{-10pt} %
    \includegraphics[width=1.1in]{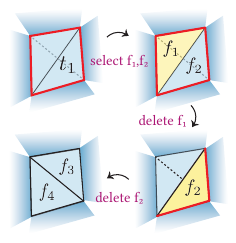}
\end{wrapfigure}
We begin by detecting all solid tetrahedral cells embedded in the triangle mesh and iteratively prune triangle pairs from each tetrahedron  without altering the overall mesh topology, \revision{processing faces in decreasing order of their area times mean edge UDF, so that faces farthest from the surface are pruned first. The inset illustrates the thinning of a single tetrahedron, with non-manifold edges highlighted in red. We then remove all \emph{open triangles}, \ie dangling triangles bounded only by open and
non-manifold edges.}

\begin{wrapfigure}[7]{r}{1.1in}
    \vspace{-16pt} %
    \hspace{-10pt} %
    \includegraphics[width=1.18in]{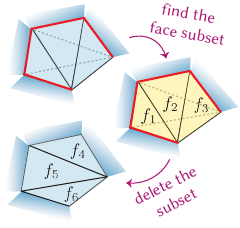}
\end{wrapfigure}
Beyond \revision{tetrahedral cells, the ambiguity of the approximate GVD also creates small polyhedral cells, where two overlapping layers meet along a cluster of non-manifold edges (see inset, the non-manifold edges are highlighted in red). We remove a subset of the incident faces such that the deletion turns every edge of the cluster manifold, and only when every deleted face is covered by the surviving ones. The full procedure is summarized in Algorithm~\ref{alg:extract}.}

The effectiveness of our results relies on both the optimization and the thinning process. As shown in \autoref{fig:method_thinning_vs_dmudf}, we conduct an ablation study in which our thinning procedure is applied as a post-processing step to both the output of DualMeshUDF~\cite{zhang2023dualmeshUDF} and our own method.

\begin{algorithm}
\revisionenv
\caption{\revision{VoroUDF optimization}}
\label{alg:opt}
\KwData{Continuous unsigned distance function $\udfunc$ with gradients $\udfgrad$}
\KwResult{Reconstructed mesh $\meshrecon$}
Initialize $N$ seeds $\{\seed_i\}_{i=1}^N$ on $\isosurface$ by uniform ray casting (Section~\ref{sec:sampling}); \\
Sample the fixed surface set $\{(\sample_m, \udfgrad_m)\}_{m=1}^M$ (Section~\ref{sec:sampling}); \\
$k \leftarrow 1$; \\
\While{\emph{relative change ratio of $\energy$} $\ge 10^{-8}$ \textbf{and} $k \le 200$}{
    Update the energy weight $\omega_k$ via Eq.~\ref{eqn:weight_schedule}; \\
    Project trial seeds onto $\isosurface$ using $\udfunc$ and $\udfgrad$; \\
    Assign each sample $\sample_m$ to its minimum-energy seed (Eq.~\ref{eqn:tangency_aware_voronoi}); \\
    Rebuild the repulsion neighbor lists $\mathcal{N}(i)$ within $5\kernel$; \\
    Evaluate $\energy$ (Eq.~\ref{eqn:energy_total}); \\
    Run L-BFGS with $\energy$ to update $N$ seeds; \\
    $k \leftarrow k + 1$; \\
}
Extract and thin the mesh from the converged seeds (Algorithm~\ref{alg:extract});
\end{algorithm}

\subsection{\revision{Discussion}}
\label{sec:discussion}

\revision{
Inspired by \cite{feng2026points} (see their Section 4), we list several less-successful formulations of our method that we considered before settling on the algorithm described in the previous sections.
}

\revision{
For example, when samples from two different nearby patches of the surface affect a single seed, the minimizer of \autoref{eqn:energy_total} can land away from the zero level-set.
To avoid this, we originally minimized (using a tailor-made ADMM optimizer) the \emph{absolute value} of the tangent residual in \autoref{eqn:dist_tan}, an \emph{$L_1$} energy in the spirit of the median attraction of the LOP operator~\cite{lipman2007lop} that should push the solution to obey the majority of samples.
Surprisingly, we found that the simpler strategy of re-projecting the minimizer of \autoref{eqn:energy_total} towards the zero level-set after every L-BFGS iteration produced more accurate results (e.g., a $30\%$ Hausdorff error improvement on \autoref{tab:exp_tab_non_manifold}) without the need to rely on more complex, purpose-made optimization schemes.

We also considered different repulsion terms before deciding on \autoref{eqn:energy_rep}. For example, as discussed in \autoref{sec:partitions}, accumulating the squared Euclidean distance $\dcov(\seed, \sample_{m}) = \lVert \seed - \sample_{m} \rVert_2^2$ over all sample-to-seed assignments yields a \emph{coverage energy}, $\energy_{cov}(\allseeds) = \sum_{m=1}^{M} \dcov(\seed_{a(m)}, \sample_{m})$, which serves as the discrete counterpart of the commonly used centroidal Voronoi tessellation (CVT) energy~\cite{liu2009centroidal}.
This energy could have served as a natural alternative to $\energy_{rep}$ in \autoref{eqn:energy_total} and forms the \emph{tangency-aware} Voronoi energy in \autoref{eqn:tangency_aware_voronoi}; however, we experimentally found that our original Gaussian repulsion energy produces far fewer topological errors and better feature recovery. We quantitatively ablate this choice in \autoref{tab:ablation_cvt_vs_repulsive_fixed_vs_decay_SALS} .
}

\begin{wrapfigure}{r}[0.15in]{2in}
    \vspace{-22pt}   %
    \hspace{-15pt} %
    \includegraphics[width=2in]{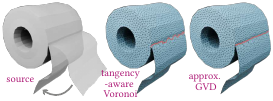}
    \vspace{-14pt}   %
\end{wrapfigure}
Finally, \revision{in our mesh extraction step, we rely on the computation of an approximate geodesic Voronoi diagram; ideally, this diagram could be used in the sample-to-seed assignment stage too, but its computational cost makes it impracticable to recompute at every L-BFGS iteration. Because of this, at the assignment stage, we settle for the tangency-aware Voronoi diagram from \autoref{eqn:tangency_aware_voronoi}.
One may reasonably wonder if this distinction matters much; i.e., whether one could use the same, faster tangency-aware Voronoi diagram for the mesh extraction step.
Experimentally, we find that the cost of the approximate GVD is well worth it in this final step, as it significantly improves our ability to accurately recover geometric features in cases where two surfaces are geodesically far yet spatially close. We show an example of this in the inset.}

%% file: tables/tab_ablation_cvt_vs_repulsive_fixed_vs_decay_SALS_reconGVD.tex
\begin{table*}[t]
\caption{\revision{
We experimented with different less-successful versions of our algorithm, which we ablate quantitatively below on 50 models from the SALS \cite{ren2025sals} dataset.
The first two rows each differ from the third in exactly one term of the seed energy: the distribution
term (coverage versus Gaussian repulsion Section~\ref{sec:discussion}) and the energy weight $\tau$. The last row keeps the same energy as the third row but replaces the
tangency-aware Voronoi partition with an approximate GVD in the mesh extraction step (Section~\ref{sec:mesh_extraction}). The \best{best}
scores are shown in bold and underlined, and the
\second{second best} scores in bold.}}
\vspace{-10pt}
\begin{center}
\resizebox{\linewidth}{!}{%
\small\setlength{\tabcolsep}{11.4pt}%
\rowcolors{2}{gray!10}{white}
\begin{tabular}{c||c|c|c|c|c|c|c}
\rowcolor{header2color}
\toprule
  Seed energy \& reconstruction &\#v & $\cd$ x$10^{-3}$~$\downarrow$ & $\hd$ x$10^{-3}$~$\downarrow$ & $\ecd$ x$10^{-2}$~$\downarrow$ &$\triangleQ$~$\uparrow$ & $\td$ ~$\downarrow$ & $\nmcd$ x$10^{-3}$~$\downarrow$ \\
\midrule
Tangent+Coverage (\autoref{eqn:tangency_aware_voronoi})
    &8.2k &\second{5.23} &\best{8.27} &1.23 &\best{0.80} &50.5 &44.25 \\
Tangent+Repulsion, fixed $\tau = 1$
    &8.2k &5.39 &9.19 &\second{0.99} &\best{0.80} &17.8 &39.56 \\
Tangent+Repulsion, decay $\tau = 0.92$
    &8.2k &5.36 &8.57 &1.04 &\best{0.80} &\second{13.6} &\second{38.29} \\
\quad use approx. GVD recon. (Ours)
    &8.2k &\best{5.02} &\second{8.47} &\best{0.59} &\best{0.80} &\best{3.0} &\best{34.55} \\
\bottomrule
\end{tabular}}
\end{center}
\label{tab:ablation_cvt_vs_repulsive_fixed_vs_decay_SALS}
\end{table*}

%% file: secs/4_results.tex
\section{Experiments}
\label{sec:experiments}

We implemented VoroUDF in Python with C++ bindings and conducted all experiments on a Linux machine with a 16-core AMD Ryzen processor and an NVIDIA RTX 3090 GPU. All 3D models in the figures are rendered using BlenderToolbox~\cite{Liu_BlenderToolbox_2018}. \textbf{We will release our code on GitHub publicly upon acceptance.}

\begin{figure}[t]
    \centering
    \includegraphics[width=\linewidth]{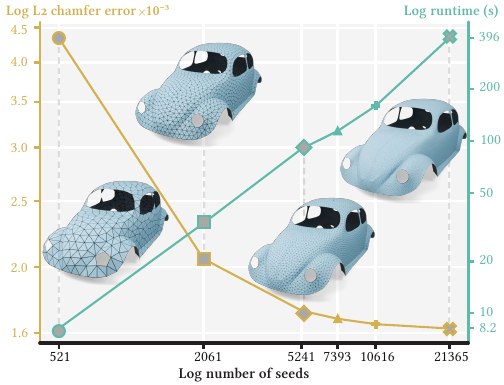}
    \vspace{-0.7cm}
    \caption{\revision{
    Our method's runtime scales near-linearly with as one increases the number of seeds and obtains an expected decrease in error.}}
    \label{fig:ablation_seeds_runtime_error}
\end{figure}

\begin{figure}[t]
    \centering
    \includegraphics[width=\linewidth]{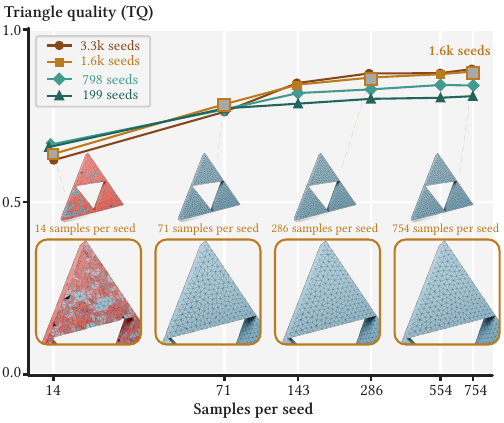}
    \vspace{-0.7cm}
    \caption{\revision{A higher number of samples per seed leads itself to higher triangle quality, with eventual diminishing returns. In practice, we fix the $286$ ratio.}}
    \label{fig:ablation_ratio_of_samples_seeds}
\end{figure}

\revision{
\paragraph{Runtime} Our method's runtime is split among three steps, illustrated in \autoref{fig:pipeline}: sampling, optimization, and mesh extraction. \autoref{fig:fig_runtime_breakdown} shows the runtime breakdown over four datasets, where the three steps take roughly 33\%, 36\%, and 31\% of the total time, respectively. We also report our method's average timings in each of the large-scale quantitative comparison tables (Tables~\ref{tab:exp_tab_non_manifold}, \ref{tab:exp_tab_SALS}, \ref{tab:exp_tab_ABC} and~\ref{tab:exp_tab_deepFashion3D}).
}

\begin{figure}[t]
    \centering
    \includegraphics[width=\linewidth]{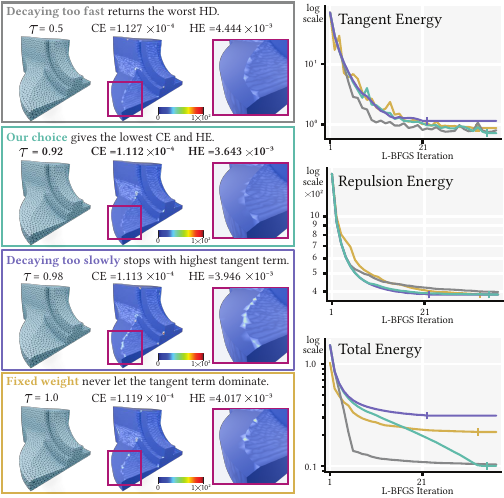}
    \vspace{-0.6cm}
    \caption{\revision{Instead of using a fixed energy weight (bottom), we rely on a weight annealing strategy with a decay parameter $\tau$ chosen to balance between over-emphasizing the tangent term in the energy too early (bottom-middle) and too late (top).}}
    \label{fig:ablation_lambda_decay}
\end{figure}

\begin{figure*}[t]
    \vspace{-10pt}
    \centering
    \includegraphics[width=\linewidth]{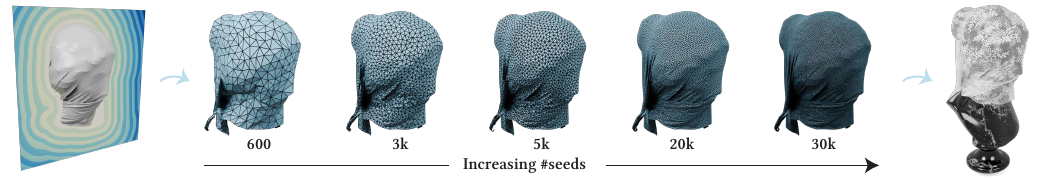}
    \vspace{-20pt}
    \caption{\revision{Level-of-detail meshing of a draped cloth (model from \citet{zhang2022progressive}). Our method extracts meshes at progressively increasing the number of seeds from $600$ to $30k$ vertices (left to right). Every level preserves the open boundaries and the folds of the drape, and the reconstructed mesh is directly usable downstream (rightmost).}}
    \label{fig:drape_sculpt}
\end{figure*}

\subsection{\revision{Parameters}}
\label{sec:parameters}

\revision{
More than any other parameter, the runtime of our method is determined by the number of output seeds $N$. Increasing it produces finer reconstructions, which is useful for surfaces with high detail. \autoref{fig:ablation_seeds_runtime_error} shows that the complexity of our method grows approximately linearly with the number of seeds, while \autoref{fig:drape_sculpt} showcases a multi-scale reconstruction example. In practice, $N$ is controlled by increasing or decreasing the number of sampled rays; since some rays miss the surface, $N$ varies slightly between results.
}
\revision{
The number of \textit{samples per seed} is another key parameter. We experimented with different values, which we ablate in \autoref{fig:ablation_ratio_of_samples_seeds}; in practice, we use $286$ samples per seed throughout our experiments.
}

\revision{
We also ablate our method's accuracy under various choices of the parameter $\omega$, which balances the repulsive term (for even distribution) against the tangent term (for feature alignment) in our energy. Rather than fixing its value, we want the optimization to first prioritize an even distribution and only later refine the seeds to align with features, which helps avoid poor local minima. We therefore borrow the weight strategy of CWF~\cite{xu2024cwf} with clamps: we keep the tangent term's weight fixed and let the repulsion weight decay, updating it once per L-BFGS iteration $k$, starting from $\omega_0 = 4$ and clamped at $\omega_{\min} = 0.3$:
\begin{equation}
\omega_k = \max\!\big(\omega_0\, \tau^{k},\ \omega_{\min}\big),
\label{eqn:weight_schedule}
\end{equation}
where we choose $\tau = 0.92$ in practice. We ablate several choices of $\tau$, including a fixed-weight strategy, in \autoref{fig:ablation_lambda_decay}, and report a quantitative comparison over an entire dataset in \autoref{tab:ablation_cvt_vs_repulsive_fixed_vs_decay_SALS}.
}

\revision{
Our method also depends on the choice of $\kernel$ (\autoref{eqn:energy_rep}), the lengthscale of the kernel used in the repulsive energy term. Intuitively, it measures how close samples need to be before they start pushing each other away; therefore, it must be a function of the surface area. As proposed by \citet{zhong2013particle}, we first estimate the surface area $A$ from the samples and fix $\kernel = 0.3 \sqrt{A / N}$ in our experiments.
}

\revision{
A final parameter of our method is the Poisson sampling radius $r$ used for sample generation~\autoref{sec:sampling}. Following the same logic as for $\kernel$, we make it depend on the estimated surface area as $\smash{r = \min(r_0, \sqrt{0.6 A / M'})}$, where $A$ is the surface area estimated from the samples, $M'$ is the number of sample candidates, and $r_0 = 0.001$; $r$ thus shrinks below $r_0$ only when the sample candidates would not otherwise fit.
}

\begin{figure}[t]
    \centering
    \includegraphics[width=\linewidth]{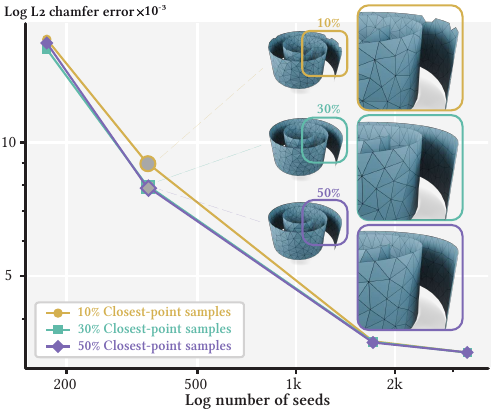}
    \vspace{-20pt}
    \caption{
    \revision{
    With a fixed samples per seed ratio, the quality of our reconstruction increases minimally but noticeably as the proportion of closest-point samples increases. In practice, we choose $30\%$ in all our experiments.}}
    \label{fig:ablation_ratio_of_grid_samples}
\end{figure}

%% file: secs/5_comparisons.tex
\section{\rewritten{Comparisons}}
\label{sec:comp}

\rewritten{Our method targets the reconstruction of lightweight meshes from modeled UDFs containing non-manifold structures, open boundaries, and sharp features. In this section, we provide comprehensive quantitative and qualitative comparisons against the most performant existing reconstruction algorithms on this task, across four datasets and along three axes: geometric fidelity, topological correctness, and triangle quality.}

\subsection{\rewritten{Methods we evaluate against}}
\label{sec:methods_compare}

We compare our method against \revision{eight existing approaches, which fall into four main categories}. The \textbf{marching-cubes-based methods} include MeshUDF~\cite{guillard2022meshudf}, CapUDF~\cite{zhou2022capudf}, GeoUDF~\cite{ren2023geoudf}, and the learning-based method NSDUDF~\cite{stella2024nsdudf}. For MeshUDF, CapUDF, and GeoUDF, we use the \rewritten{\textit{standalone mesh extraction modules}} provided in their released code. Since CapUDF's extraction duplicates vertices per grid cell, we apply the same post-processing step used in MeshUDF to remove duplicate vertices for consistency. For NSDUDF, we evaluate the publicly released model trained on the ABC dataset across all benchmarks. We also compare against two \textbf{double-covering-based methods}, DCUDF~\cite{hou2023dcudf} and DCUDF2~\cite{chen2025dcudf2}, one \textbf{dual-contouring-based method}, DualMeshUDF~\cite{zhang2023dualmeshUDF}, \revision{and one \textbf{material-interface-based method}, MIND~\cite{chen2026mind}, which assigns multiple labels to a volumetric grid and extracts the interfaces between them, making it the baseline closest to ours in its explicit handling of non-manifold structures}.

\subsection{Evaluation Metrics}
\label{sec:eval_metrics}

To evaluate the quality of the reconstructed meshes, we adopt several standard metrics. For mesh regularity, we report \textit{Triangle Quality} (\triangleQ)~\cite{frey1999surface} and \textit{Topology Error} ($\td$). To assess geometric fidelity between the reconstructed surface and the ground truth, we use the \revision{$L_2$} \textit{Chamfer Error} ($\cd$) and the \textit{Hausdorff Error} ($\hd$). To evaluate sharp feature reconstruction, we report the \revision{$L_2$} \textit{Edge Chamfer Error} ($\ecd$)~\cite{chen2021neuralmc}. For non-manifold regions, we introduce the \textit{Non-Manifold Chamfer Error} ($\nmcd$) to measure reconstruction accuracy. \rewritten{We define each metric below.}

\input{tables/tab_exp_SALS.tex}

\paragraph{Triangle Quality (\triangleQ)}
\rewritten{We use the \triangleQ~indicator~\cite{frey1999surface}, where a value closer to $1.0$ indicates a nearly equilateral triangle:
\begin{equation}
\triangleQ(t) = \frac{6}{\sqrt{3}} \cdot \frac{S_t}{p_t \cdot h_t},
\end{equation}
where $S_t$, $p_t$, and $h_t$ denote the area, half-perimeter, and longest edge length of triangle $t$, respectively; we report the average over all triangles of the reconstructed mesh.}

\paragraph{$L_2$ Chamfer Error ($\cd$)}
\rewritten{The Chamfer error measures the overall geometric discrepancy between the reconstructed mesh and the ground truth. We sample $n = 10^6$ points uniformly at random on each of the two meshes, yielding point sets $\mathcal{P}$ (reconstruction) and $\mathcal{Q}$ (ground truth), and match every sample to its nearest neighbor in the other set. We report the symmetric $L_2$ Chamfer error, which aggregates the squared nearest-neighbor distances into a root mean square on each side:
\begin{equation}
\cd(\mathcal{P}, \mathcal{Q}) =
\sqrt{\frac{1}{n} \sum_{\mathbf{p} \in \mathcal{P}} \min_{\mathbf{q} \in \mathcal{Q}} \left\| \mathbf{p} - \mathbf{q} \right\|_2^2}
\;+\;
\sqrt{\frac{1}{n} \sum_{\mathbf{q} \in \mathcal{Q}} \min_{\mathbf{p} \in \mathcal{P}} \left\| \mathbf{q} - \mathbf{p} \right\|_2^2}.
\label{eqn:cd_l2}
\end{equation}
} %

\paragraph{$L_2$ Edge Chamfer Error ($\ecd$)}
\rewritten{To assess the accuracy of sharp feature reconstruction, we adopt the \emph{edge} Chamfer distance of~\citet{chen2021neuralmc} and follow its reference implementation. We draw $2.5 \times 10^5$ surface samples, each carrying the normal of its face, on both the ground-truth mesh and the reconstruction. A sample is classified as an \emph{edge point} if some other sample within a radius of $0.005$ has a strongly deviating normal, $|\mathbf{n}_i^{\top} \mathbf{n}_j| < 0.2$ (a dihedral angle of more than $78^\circ$); this selects the samples straddling creases and sharp corners. Let $\mathcal{P}_e$ and $\mathcal{Q}_e$ denote the edge points of the reconstruction and of the ground truth. $\ecd$ is their symmetric $L_2$ Chamfer error, the sum of the two mean squared nearest-neighbor distances,
\begin{equation}
\ecd = \frac{1}{|\mathcal{P}_e|} \sum_{\mathbf{p} \in \mathcal{P}_e} \min_{\mathbf{q} \in \mathcal{Q}_e} \left\| \mathbf{p} - \mathbf{q} \right\|_2^2
\;+\;
\frac{1}{|\mathcal{Q}_e|} \sum_{\mathbf{q} \in \mathcal{Q}_e} \min_{\mathbf{p} \in \mathcal{P}_e} \left\| \mathbf{q} - \mathbf{p} \right\|_2^2,
\label{eqn:ecd_l2}
\end{equation}
}

\paragraph{Hausdorff Error ($\hd$)}
To detect missing or redundant geometry (\eg a hole or a floating piece), we report the Hausdorff error: $\hd^1$ is the one-sided Hausdorff distance from the ground-truth surface to the reconstructed mesh, $\hd^2$ the distance in the reverse direction, and $\hd = \max(\hd^1, \hd^2)$. All Hausdorff distances are normalized by the diagonal length of the model's bounding box.

\begin{figure}[t]
    \centering
    \includegraphics[width=\linewidth]{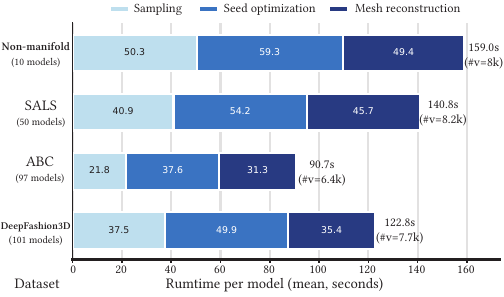}
    \vspace{-0.7cm}
    \caption{\revision{Runtime breakdown of our pipeline over four datasets, split into the three stages of the pipeline: sampling (Section~\ref{sec:sampling}), seed optimization (Section~\ref{sec:energy}), and mesh reconstruction (Section~\ref{sec:mesh_extraction}); the total and the mean output vertex count ($\#v$) are printed at the end of each bar. All timings are CPU-only on a reported Linux machine.}}
    \label{fig:fig_runtime_breakdown}
\end{figure}

\begin{figure*}[t!]
    \centering
    \includegraphics[width=\linewidth]{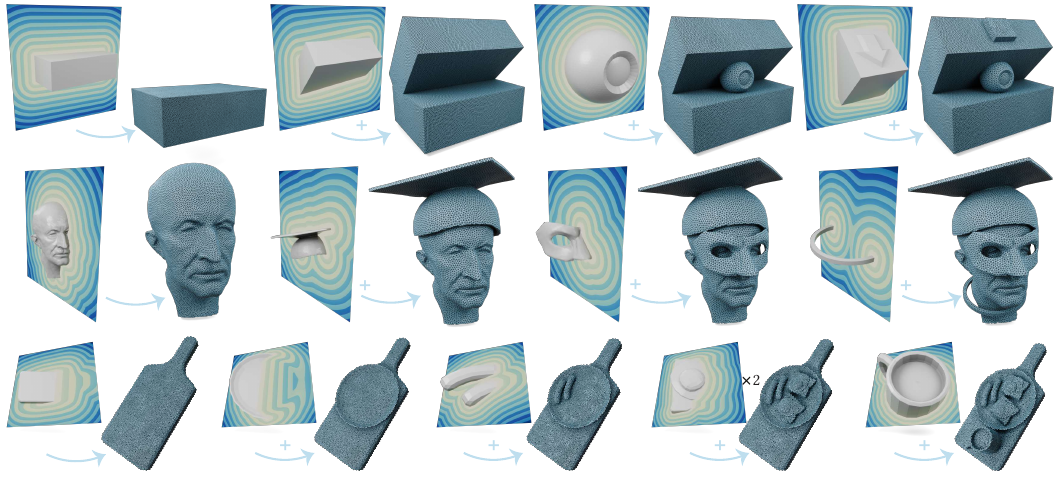}
    \vspace{4pt}
    \includegraphics[width=\linewidth]{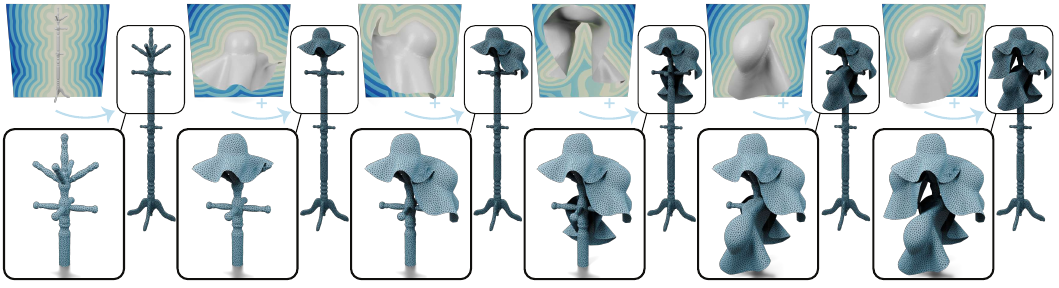}
    \vspace{-10pt}
    \caption{\revision{Incremental modeling with UDFs that a new part is merged into the unsigned field and the accumulated shape is re-meshed by our method after every edit. The coat rack models are from \citet{zhang2025progressive}.}}
    \label{fig:modeling}
\end{figure*}

\paragraph{Topology Error ($\td$)}
We use the \textit{Euler characteristic} as a measure of mesh topology and report the absolute difference between the Euler characteristic of the source UDF isosurface and that of the reconstructed mesh. This metric reflects the topological discrepancy between the reconstruction and the ground truth.

\paragraph{Non-Manifold Chamfer Error ($\nmcd$)}
\rewritten{The ground-truth shapes of our non-manifold benchmarks are assembled from separate parts that are not welded into a single mesh, so their non-manifold curves are not explicit edges of the ground-truth mesh; we therefore define the ground-truth non-manifold set geometrically. We split the ground-truth mesh into its connected components, draw $2 \times 10^5$ surface samples in total, distributed over the components in proportion to their area, and keep every sample that lies within a distance of $0.005$ (in the $[-1,1]^3$ frame of the normalized models) of a sample on a \emph{different} component, \ie the points where the sheets meet or pass through each other. On the reconstruction, we simply detect the non-manifold edges (shared by three or more faces) and draw $5 \times 10^4$ samples along them. $\nmcd$ is the symmetric $L_2$ Chamfer error between the two sets, computed as in \autoref{eqn:ecd_l2}, and quantifies both how accurately the non-manifold structures are placed and whether spurious ones are introduced. Reconstructions with no non-manifold edge detected have no such samples, and we report their $\nmcd$ as $\text{Inf}$.}

\begin{figure*}[t!]
    \centering
    \includegraphics[width=\linewidth]{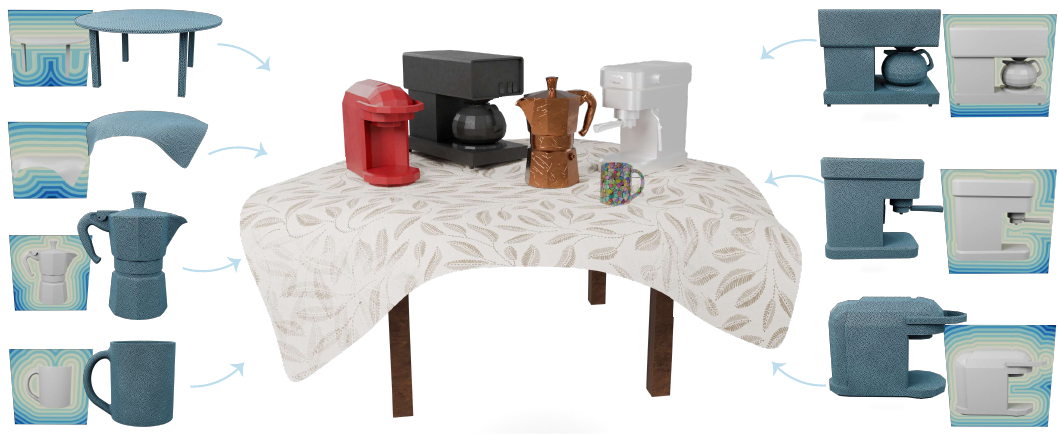}
    \vspace{-10pt}
    \caption{\revision{Composing a scene from independently assets. Each object, from the table, the draped tablecloth, and a mug to the four coffee makers, is represented by its own unsigned distance field and meshed separately by our method, covering open boundaries, thin handles, and sharp CAD-like features. The reconstructed meshes are then assembled into a single scene (center), with no per-asset tuning.}}
    \label{fig:table}
\end{figure*}

\subsection{\rewritten{Datasets}}

\rewritten{To comprehensively evaluate our method against the selected baselines, we assemble four datasets, each isolating one class of inputs that our method is designed to handle.}

One of our method's stated goals is to recover shapes with \emph{non-manifold} structures, which we benchmark on two synthetic datasets. The first is a dataset of 10 non-manifold models (see inset) that we construct ourselves: we select a mesh from the ABC dataset~\cite{koch2019abc}, cut it with a plane, and merge the two parts back 
\begin{wrapfigure}[7]{r}{1.6in}
    \vspace{-5pt} %
    \hspace{-15pt} %
    \includegraphics[width=1.7in]{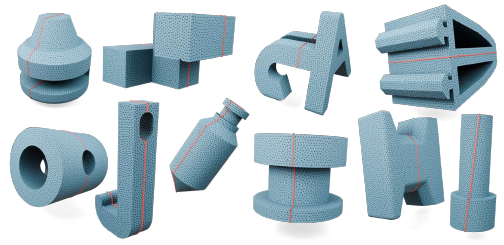}
\end{wrapfigure}
with a narrow overlapping band (see \autoref{fig:more_nonmanifold_d7_NW1}). By construction, this dataset comes with reliable ground truth for its non-manifold regions. We densely sample points on each of the two cut meshes and identify those near the opposing part (see red points in \autoref{fig:more_nonmanifold_d7_NW1}). These red samples approximate the ground-truth non-manifold edges of the merged mesh, enabling us to compute the Non-Manifold Chamfer Error (\nmcd), detail see Section~\ref{sec:eval_metrics}. Results on this dataset are presented in \autoref{tab:exp_tab_non_manifold} and \autoref{fig:more_nonmanifold_d7_NW1}. The second is the dataset introduced by \citet{ren2025sals}, which combines two ABC shapes to produce self-intersections; results are reported in \autoref{tab:exp_tab_SALS} and \autoref{fig:more_SALS_nm1_d7}.

\rewritten{
To evaluate performance on \emph{manifold, closed surfaces with sharp features}, we use the same subset of 100 shapes from the ABC dataset \cite{koch2019abc} adopted by \citet{xu2024cwf}. Results on this dataset are reported in \autoref{tab:exp_tab_ABC} and \autoref{fig:more_abc_d7_NM1_1}.
Second, we consider \emph{open surfaces}, represented by 100 garment models from the DeepFashion3D dataset~\cite{zhu2020deepfashion}. Results are reported in \autoref{tab:exp_tab_deepFashion3D} and exemplified in \autoref{fig:more_deepFashion3D_d7_NM1_1}.
}

\subsection{\rewritten{Quantitative Results}}
\label{sec:quatitative_results}

\rewritten{Our goal is to output \emph{lightweight} meshes (i.e., below $10{,}000$ vertices) that can be directly used in interactive and real-time downstream applications. We therefore begin by evaluating all methods at these low-to-medium output resolutions, reported in the rows, with resolution $2^5$ and $2^7$, of Tables~\ref{tab:exp_tab_non_manifold}, \ref{tab:exp_tab_SALS}, \ref{tab:exp_tab_ABC} and~\ref{tab:exp_tab_deepFashion3D}.}

\rewritten{A critical property of any reconstruction method at this resolution is its \emph{accuracy} in approximating the true zero level set of the UDF. At low resolution ($2^5$ of first rows of Tables~\ref{tab:exp_tab_non_manifold}, \ref{tab:exp_tab_SALS} and~\ref{tab:exp_tab_ABC}) , our method (last row of each table) outperforms all prior methods across every accuracy metric: our Chamfer and Hausdorff errors are significantly lower (in some cases by an order of magnitude). As shown in Tables~\ref{tab:exp_tab_non_manifold}, \ref{tab:exp_tab_SALS} and~\ref{tab:exp_tab_ABC}, our method also excels at preserving sharp features, yielding noticeably lower edge Chamfer errors ($\ecd$). Similarly, our method accurately recovers non-manifold structures: most competing methods are by construction unable to produce non-manifold edges at all, and our algorithm outperforms even those that can (e.g., DualMeshUDF) severalfold in non-manifold Chamfer error ($\nmcd$).}

\rewritten{To be directly usable in the applications we target, our output meshes must be not only lightweight and accurate but also \emph{high-quality}: valid inputs for geometry processing tasks such as texturing, deformation, and physical simulation. Our method encourages this quality through the repulsive term of our energy (\autoref{eqn:energy_rep}), and its effect is clearly reflected in the results: our method consistently produces the meshes with the highest triangle quality ($\triangleQ$) and lowest topology error ($\td$) across all four tables (Tables~\ref{tab:exp_tab_non_manifold}, \ref{tab:exp_tab_SALS}, and~\ref{tab:exp_tab_ABC}).}

\rewritten{In sum, these results show that our algorithm sets a new state of the art for producing lightweight, high-quality, accurate mesh reconstructions from unsigned distance functions. We push this advantage further by comparing our meshes ($6$--$8$k vertices) against the output of existing methods run \emph{at medium resolution} (tens to hundreds of thousands of vertices with resolution $2^7$), shown in the lower half of the same tables. Even at these much higher output resolutions, competing reconstructions are merely comparable in terms of geometric accuracy measured by Chamfer $\ecd$ and Hausdorff $\hd$ error, but still inferior to ours in triangle quality $\triangleQ$ and topology error $\td$, while using an order of magnitude more vertices.}

\begin{figure*}[t]
    \centering
    \includegraphics[width=\linewidth]{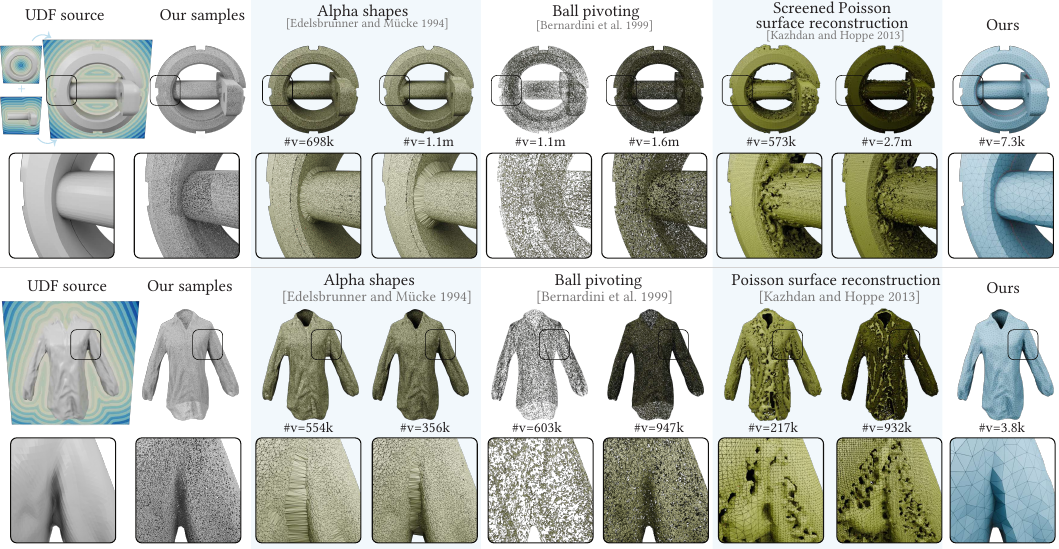}
    \vspace{-0.7cm}
    \caption{\revision{
    Our method relies on a dense sampling of the surface, computed using the UDF and its gradient. By fully exploiting the information contained in the UDF, it produces more accurate reconstructions with far fewer vertices than if one input the same samples to off-the-shelf point cloud reconstruction methods.
    The lack of consistently-oriented gradients causes screen Poisson reconstruction to produce noticeable artifacts.}}
    \label{fig:comp_pointcloud}
\end{figure*}

\subsection{\rewritten{Qualitative Results}}
\label{sec:qualitative_results}

\rewritten{While the numerical results in Tables~\ref{tab:exp_tab_non_manifold}, \ref{tab:exp_tab_SALS}, and~\ref{tab:exp_tab_ABC} demonstrate our method's superiority at its stated task, we now turn to individual examples to show how this superiority manifests and to build intuition as to why. In general, as shown in Figures~\ref{fig:teaser} and \ref{fig:rw_nonmanifold}, our method is the only one able to reconstruct the most challenging non-manifold structures at the low-to-medium resolutions we consider.}

\rewritten{We investigate the reasons behind this with targeted comparisons. In \autoref{fig:method_vs_dcudf}, we illustrate why double-cover-based methods (which mesh a double-layered offset surface and then shrink or cut it open) are structurally unable to produce non-manifold features: their cutting stage can only yield manifold patches. In \autoref{fig:method_vs_mc}, we compare our method against marching-cubes-based approaches such as MeshUDF~\cite{guillard2022meshudf} and CapUDF~\cite{zhou2022capudf}. Inheriting the fixed grid of marching cubes, these methods produce meshes whose vertices and edges do not align with the shape's sharp features; while this artifact fades at higher resolutions, it renders the meshes clearly unsuitable at the lightweight resolutions we target.}

\rewritten{Finally, in \autoref{fig:method_vs_dmudf}, we illustrate why contouring-based methods such as DualMeshUDF~\cite{zhang2023dualmeshUDF} do not fare better in our tests: at these lower resolutions, a single grid cell frequently spans several sheets of the surface, forcing the method to output a single vertex per cell leads to topological noise and holes.}

\begin{figure}[t]
    \centering
    \includegraphics[width=\linewidth]{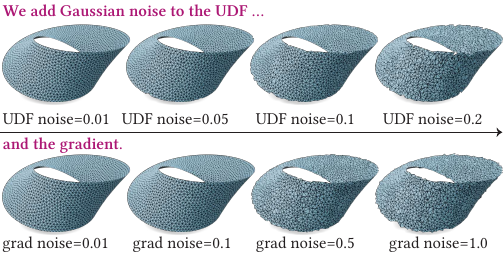}
    \vspace{-0.7cm}
    \caption{\revision{Noise robustness on a M\"obius band. Top: relative distance noise, $\tilde{\udfunc} = \udfunc \cdot (1+\varepsilon)$ with $\varepsilon \sim \mathcal{N}(0, \sigma^2)$~\cite{stella2024nsdudf}, at $\sigma$ up to $0.2$. Bottom: gradient noise, added per query and renormalized, at $\sigma$ up to $1.0$.}}
    \label{fig:noise}
\end{figure}

\subsection{\revision{Comparison to point cloud reconstruction}}
\label{sec:comp_to_point_cloud}

\revision{Our method uses a large number of surface point samples to produce its mesh output; thus, in a sense, it relies on a point cloud as an intermediate shape representation. A natural question is if one could skip our mesh optimization and extraction steps altogether and replace them with an existing point cloud reconstruction method such as Alpha Shapes~\cite{edelsbrunner1994alpha}, Ball Pivoting~\cite{bernardini1999ball}, and Screened Poisson Reconstruction~\cite{kazhdan2013screened}. We explore this question in \autoref{fig:comp_pointcloud}. By exploiting not only the samples themselves but also the information carried by the underlying UDF, our method avoids the artifacts of purely point-based methods while producing significantly more compact meshes. Methods like Poisson Surface Reconstruction \cite{kazhdan2006poisson,kazhdan2013screened} fail at this task altogether, as UDF gradients are not consistently oriented.}

\revision{So far, we have assumed that the input to our method is an exact UDF induced by a two-dimensional surface. One may wonder, however, whether our method can also reconstruct surfaces \emph{directly from point cloud inputs}, by defining the input field as the distance to the closest point in the cloud. We experiment with this in \autoref{fig:comp_pointcloud_inputs} which shows our algorithm's surprisingly high fidelity and sharp feature reconstruction with a significantly (three or four orders of magnitude) smaller output vertex budget.}

%% file: tables/tab_exp_SALS.tex
\begin{table*}[t]
\caption{\revision{Quantitative comparison on 50 intersecting models of
the SALS dataset~\cite{ren2025sals} against eight state-of-the-art methods, each evaluated at
resolutions $2^5$ and $2^7$; our Voronoi-based
method has no grid resolution, so its single run is the last row. The
\best{best} scores are shown in bold and underlined, and the
\second{second best} scores in bold. For prior methods
produce outputs without non-manifold structure, we mark $\nmcd$ as Inf.}}
\vspace{-10pt}
\begin{center}
\resizebox{\linewidth}{!}{%
\small\setlength{\tabcolsep}{9.7pt}%
\rowcolors{2}{gray!10}{white}
\begin{tabular}{c|c||c|c|c|c|c|c|c|c}
\rowcolor{header2color}
\toprule
  Method &Res. &\#v & $\cd$ x$10^{-3}$~$\downarrow$ & $\hd$ x$10^{-3}$~$\downarrow$ & $\ecd$ x$10^{-2}$~$\downarrow$ &$\triangleQ$~$\uparrow$ & $\td$ ~$\downarrow$ & $\nmcd$ x$10^{-3}$~$\downarrow$ & Time (s)\\
\midrule
MeshUDF &$2^5$
    &6.6k &70.60 &42.76 &7.04 &0.64 &170.4 &Inf &28.5 \\
CapUDF &$2^5$
    &23.5k &77.69 &93.34 &4.74 &0.57 &5826.1 &Inf &29.1 \\
GeoUDF &$2^5$
    &4.1k &28.26 &44.96 &12.50 &0.65 &101.8 &Inf &29.0 \\
NSDUDF &$2^5$
    &3.4k &21.30 &34.89 &21.24 &0.64 &98.2 &Inf &4.1 \\
DCUDF &$2^5$
    &6.6k &103.20 &44.53 &26.45 &0.72 &17.8 &Inf &866.8 \\
DCUDF2 &$2^5$
    &8.6k &86.32 &47.41 &35.46 &0.71 &22.9 &Inf &1841.0 \\
DualMeshUDF &$2^5$
    &5.1k &14.56 &38.87 &1.41 &0.52 &822.0 &146.63 &11.9 \\
    \midrule
    \midrule
MeshUDF &$2^7$
    &118.9k &16.19 &11.97 &6.02 &0.66 &1220.5 &Inf &94.8 \\
CapUDF &$2^7$
    &471.2k &11.59 &19.37 &4.19 &0.59 &118k &Inf &1822.0 \\
GeoUDF &$2^7$
    &71.4k &6.07 &12.93 &7.20 &0.67 &974.2 &Inf &520.9 \\
NSDUDF &$2^7$
    &60.4k &\second{5.65} &\second{8.22} &12.97 &0.65 &536.5 &Inf &17.2 \\
DCUDF &$2^7$
    &154.7k &20.21 &8.76 &48.73 &0.72 &32.9 &Inf &931.2 \\
DCUDF2 &$2^7$
    &228.2k &12.92 &\best{7.93} &56.35 &0.71 &\second{9.3} &Inf &3715.8 \\
DualMeshUDF &$2^7$
    &88.3k &5.99 &10.12 &\best{0.52} &0.59 &16k &162.64 &165.8 \\
MIND &$2^7$
    &63.8k &26.07 &117.00 &14.46 &\second{0.75} &246.5 &\second{105.79} &588.4 \\
    \midrule
    \midrule
Ours &/
    &8.2k &\best{5.02} &8.47 &\second{0.59} &\best{0.80} &\best{3.0} &\best{34.55} &140.8 \\
\bottomrule
\end{tabular}}
\end{center}
\label{tab:exp_tab_SALS}
\end{table*}

%% file: secs/6_application.tex
\section{Applications}
\label{sec:application}

\revision{
UDFs are increasingly common in artistic modeling applications.
Like polygonal meshes, they can represent shapes with open boundaries, non-manifold edges and sharp features. Unlike meshes, however, shapes described as UDFs can be easily combined with one another in elaborate modeling frameworks without the need for invasive mesh surgery; for example, by taking the pointwise minimum of the two UDFs to represent their union.

In \autoref{fig:modeling}, we prototype such iterative 3D modeling applications, in which each component of the object is modeled independently and its mesh extracted with our method. Isotropic and in the thousands of vertices, the output meshes produced lead themselves to efficient texturing, rendering and even tetrahedralization and physical simulation.
A similar application is highlighted in \autoref{fig:table}, in which each asset in a complex scenes is reconstructed from a UDF independently and composited together.

Beyond combining shapes, a common workflow for 3D artists is to \emph{sculpt} elaborate shapes by \emph{sweeping} a brush along a trajectory (see, e.g., \cite{adobe_substance_modeler}).
As we showcase in \autoref{fig:swept_envelope}, when combined with our method, UDFs may be an underexplored tool for the computation of such sweeps. While their \emph{signed} counterparts have been used before to compute swept volumes \cite{sellan2021swept}, recent work has explored the shortcomings of this strategy, as the output of sweeping an SDF brush is no longer an SDF but a \emph{pseudo-}SDF with reduced mathematical properties \cite{marschner2023constructive}.
UDFs present no such shortcoming, and may provide a useful tool for swept volume computation in the future.
}

\begin{figure}[t]
    \vspace{-0.3cm}
    \centering
    \includegraphics[width=\linewidth]{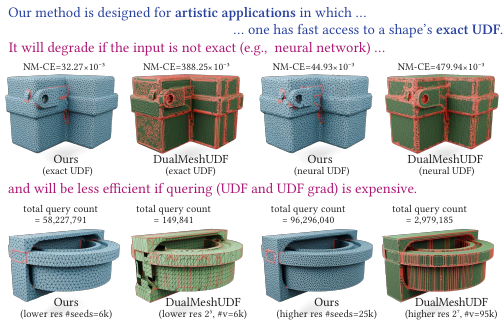}
    \vspace{-0.7cm}
    \caption{\revision{Our method is designed for applications with \emph{fast} access to a shape's \emph{exact} UDF and gradients, and may be less suitable for cases in which the UDF is approximated by a large neural network where each query requires a costly forward pass. In this case, purpose-made algorithms like DualMeshUDF \cite{zhang2023dualmeshUDF} may be preferable.}}
    \vspace{-0.5cm}
    \label{fig:limitation_neural_query}
\end{figure}

\begin{figure}[t]
    \centering
    \includegraphics[width=\linewidth]{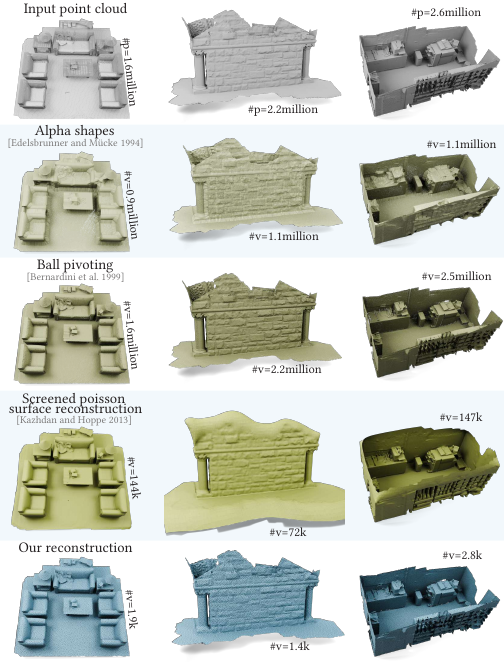}
    \vspace{-0.7cm}
    \caption{\revision{
    An input point cloud can be used to compute unsigned distances from, leading to a functional input that can be consumed by our method to produce lightweight reconstructions with open surfaces from real scan data \cite{zhou2013dense}.}}
    \vspace{-0.5cm}
    \label{fig:comp_pointcloud_inputs}
\end{figure}

%% file: secs/7_conclusion.tex
\section{Limitations \& Conclusion}
We have introduced VoroUDF, a surface reconstruction algorithm that generates high-quality triangle meshes from unsigned distance fields. Unlike all prior work on this task, our results preserve faithful non-manifold structures, sharp features, and open surfaces. Our method leverages a Voronoi-based partitioning framework, \revision{combined with a tangent energy and a Gaussian repulsive energy}, to achieve state-of-the-art performance across a wide range of input scenarios.

\begin{figure*}[t!]
    \centering
    \includegraphics[width=\linewidth]{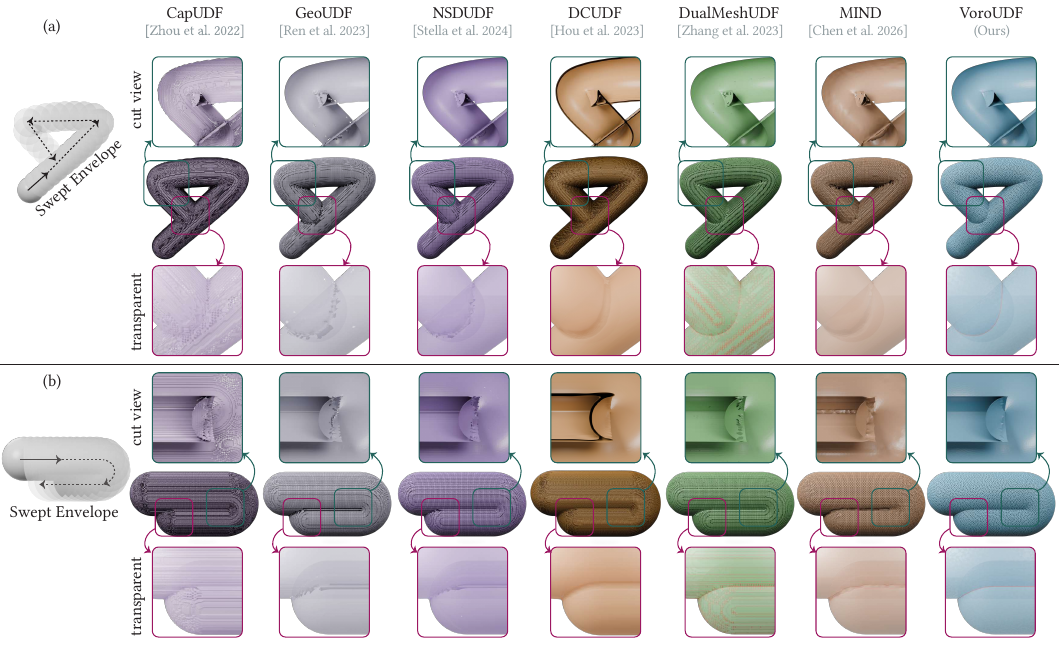}
    \vspace{-20pt}
    \caption{\revision{Two shapes are swept along two distinct paths and reconstructed using our method and our considered baselines. Prior methods produce spurious connections or double layers, while ours reconstructs the non-manifoldness and interior sheets cleanly.}}
    \vspace{-0.5cm}
    \label{fig:swept_envelope}
\end{figure*}

Our Voronoi-based spatial partitioning eliminates the need for subdividing voxel grids and is better suited for preserving thin structures. However, it requires seed points to be initialized near such features; without adequate initialization, small-scale geometry may be missed or poorly reconstructed. In future work, we plan to explore adaptive or data-driven seed initialization strategies to ensure sufficient sampling around thin or intricate structures.

\begin{wrapfigure}[7]{r}{1.75in}
    \vspace{-15pt} %
    \hspace{-16pt} %
    \includegraphics[width=1.95in]{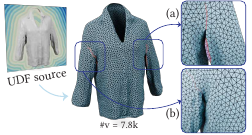}
\end{wrapfigure}
Across every benchmark we considered, our method recovers nonmanifold junctions, sharp features and open boundaries that no prior UDF mesher captures, and it does so with an order of magnitude fewer vertices. \revision{It offers, however, no formal guarantee on the topology of its output: neither the seed optimization nor the dual extraction guarantees that the result is free of holes, and our evidence for topological fidelity is purely empirical. The failure modes we observe concentrate in one configuration, where two layers of the surface are at a very narrow distance comparable to the seed spacing and the samples of both layers fall into the neighborhood of a single seed. The outcome then depends on how the seeds settle, and both possibilities occur, sometimes within the same model (see inset): the two layers may be merged into a single sheet, losing the gap between them (inset (b), highlighted single layer sheet in purple), or they may stay separate but be stitched together by a few non-manifold edges (inset (a), highlighted in red), even when the input surface is a true manifold. Incorporating topological validation or correction, for instance, re-examinate the extracted dual against the pruned sample graph, or re-seeding locally where a cell straddles two layers, is a potential direction for future work.}

\revision{
Motivated by the application of constructive modeling with simple primitives, our algorithm is designed to operate on \emph{exact} SDFs that can be \emph{efficiently queried}:
in particular, our algorithm relies on the ablity to project points onto the surface by following the UDF gradient by an amount equal to its magnitude, $\seed \rightarrow \seed - |\udfunc(x)|\nabla \udfunc(x)$.
If one swaps this direct projection for a (more costly) limited number of gradient descent steps, our algorightm can become robust to small to moderate amounts of noise (see \autoref{fig:noise}).
}

\revision{
However, even so, the ability of our method to recover exact sharp and non-manifold features will degrades as the noise in the UDF signal increases, as in the case of the neural distance field shown in \autoref{fig:limitation_neural_query}.
Indeed, further improvements will be needed to make our method suitable for the \emph{deep} UDFs that are popular in 3D generative applications \cite{zhou2024udiff,liu2024dreamudf,zhou2026udfstudio}. As shown in \autoref{fig:limitation_neural_query} (bottom), our method also relies on a high number of UDF queries, potentially leading to a high computational cost if each query necessitates a full forward pass of an elaborate neural architecture.
}

Our work focuses on reconstructing UDFs with complex geometries and topologies. With it, we aim to contribute to a growing body of research that helps bridge the gap between the flexible modeling practices of 3D artists and the geometric quality and fidelity requirements of downstream applications. 
We also seek to help connect the 3D generative deep learning and geometry processing communities, enabling generated shapes to be used in real-world tasks like rendering, physical simulation, and animation.

\begin{algorithm}
\revisionenv
\caption{\revision{VoroUDF mesh extraction and thinning (Section~\ref{sec:mesh_extraction})}}
\label{alg:extract}
\KwData{UDF $\udfunc$ with gradients $\udfgrad$; converged seeds $\{\seed_i\}_{i=1}^N$; \\
Poisson ray-cast samples $\{(\sample_m, \udfgrad_m)\}$}
\KwResult{Reconstructed mesh $\meshrecon$}
Build a kNN graph over the samples and the seeds\\
prune graph edges with $|\udfunc(\text{midpoint})| > \thredudf = 0.1$ or misaligned with $\udfgrad$ at both sample endpoints ($> \thredudfgrad = 0.3$); \\
Label every sample with its geodesically closest seed: multi-source Dijkstra over the graph from the seed; \\
\ForEach{sample $\sample_m$}{
    $L \leftarrow$ labels in the $1$-ring of $\sample_m$; \\
    $|L| = 3$: add triangle; \\
    $|L| = 4$: add the 4 triangles of the tetrahedron; \\
    $5 \le |L| \le 8$: fan-triangulate the polygon of seeds; \\
    else discard; \\
}
\tcp{Thinning}
Record every 4 seeds whose 4 triangles are all present as a tetrahedron; \\
Query $\udfunc$ at the midpoints of all cell edges; \\
\While{a tetrahedral cell remains}{
    Pop the free face with largest area $\times$ mean edge UDF; \\
    Select a face pair $(f_a, f_b)$ of its tet; \\
    Collapse the simple pairs (tet, $f_a$) and ($f_b$, freed edge); \\
}
Delete open triangles until none remain; \\
\ForEach{cluster of non-manifold edges}{
    Find a face subset whose deletion turns every adjacent edges manifold; \\
    delete it if every face is covered by the surviving faces; \\
}
\end{algorithm}

%% file: secs/10_app.tex
\input{tables/tab_exp_non_manifold.tex}
\input{tables/tab_exp_ABC.tex}
\input{tables/tab_exp_deepFashion3D.tex}

\begin{figure*}
    \centering
    \includegraphics[width=0.98\linewidth]{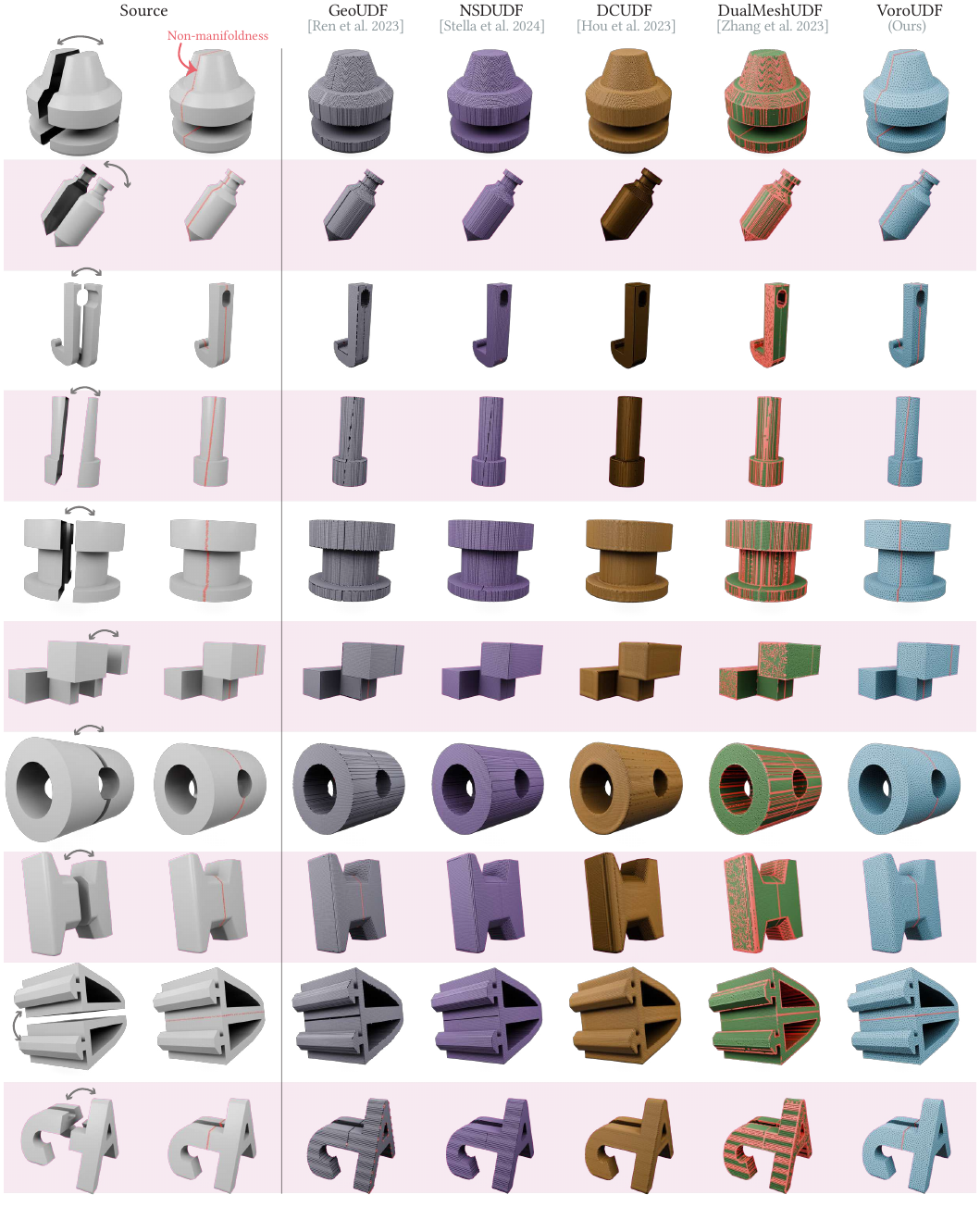}
    \vspace{-10pt}
    \caption{\rewritten{Qualitative comparison on 10 synthetic models with non-manifold edges, evaluated against seven state-of-the-art methods at higher resolution ($2^7$). Non-manifold edges are highlighted in red.}}
    \label{fig:more_nonmanifold_d7_NW1}
\end{figure*}

\begin{figure*}
    \centering
    \includegraphics[width=0.98\linewidth]{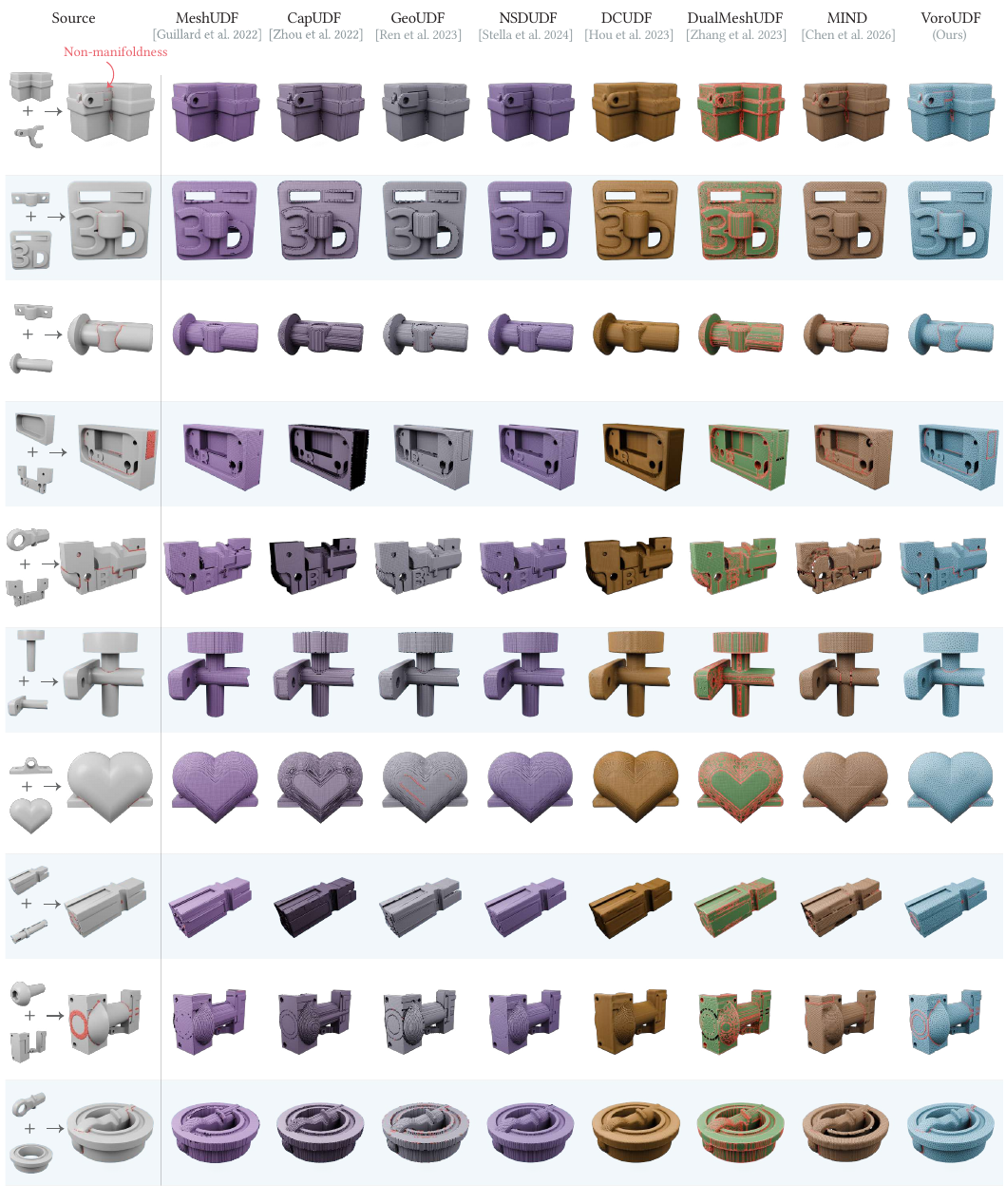}
    \vspace{-10pt}
    \caption{\revision{Qualitative comparison on 50 SALS~\cite{ren2025sals} models (part 1) merging two ABC models with nonmanifoldness, evaluated against seven state-of-the-art methods at resolution ($2^7$). Non-manifold edges are highlighted in red.}}
    \label{fig:more_SALS_nm1_d7}
\end{figure*}

\begin{figure*}
    \centering
    \includegraphics[width=0.98\linewidth]{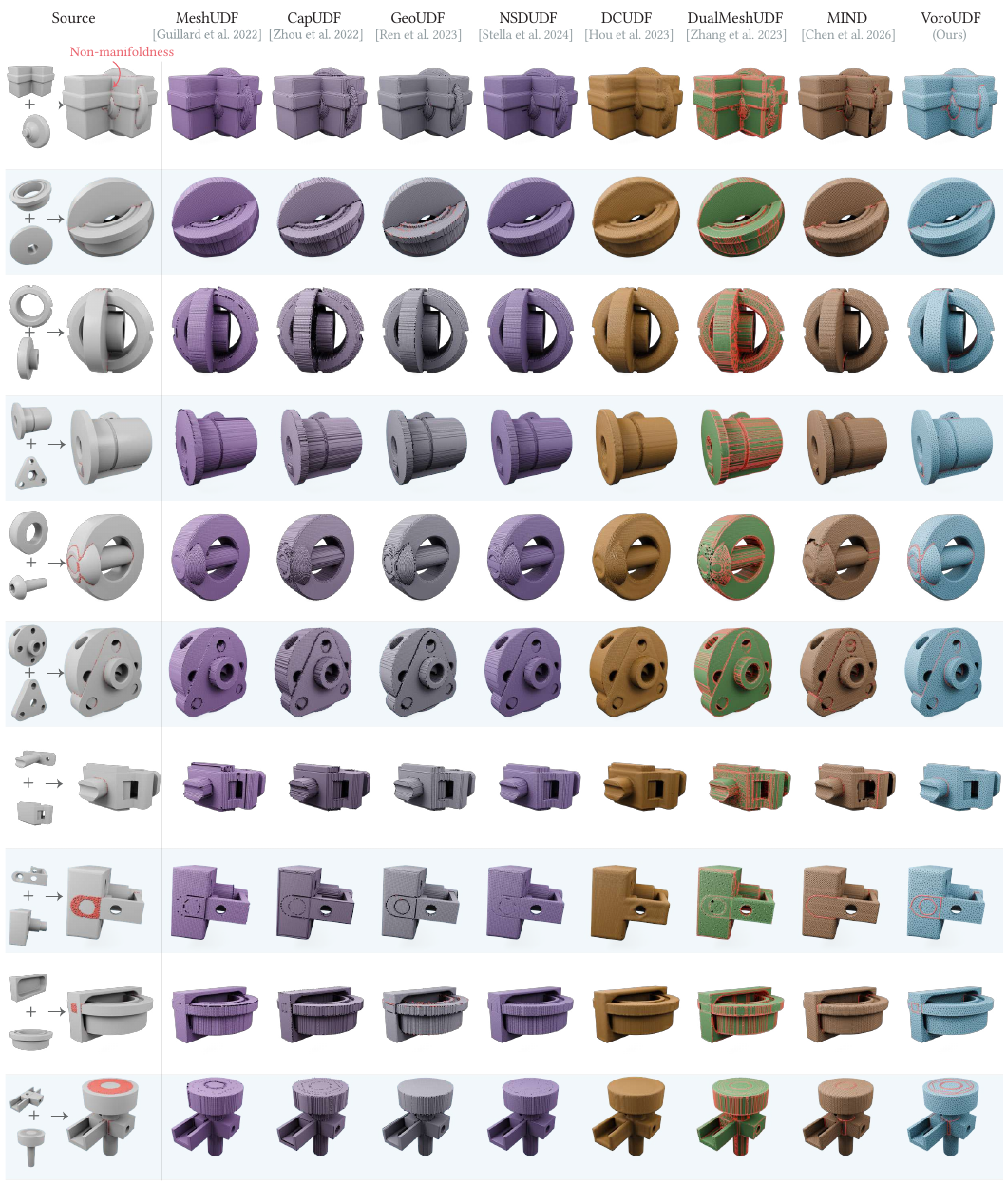}
    \vspace{-10pt}
    \caption{\revision{Qualitative comparison on 50 SALS~\cite{ren2025sals} models (part 2) merging two ABC models with nonmanifoldness, evaluated against seven state-of-the-art methods at resolution ($2^7$). Non-manifold edges are highlighted in red.}}
    \label{fig:more_SALS_nm1_d7}
\end{figure*}

\begin{figure*}
    \centering
    \includegraphics[width=0.96\linewidth]{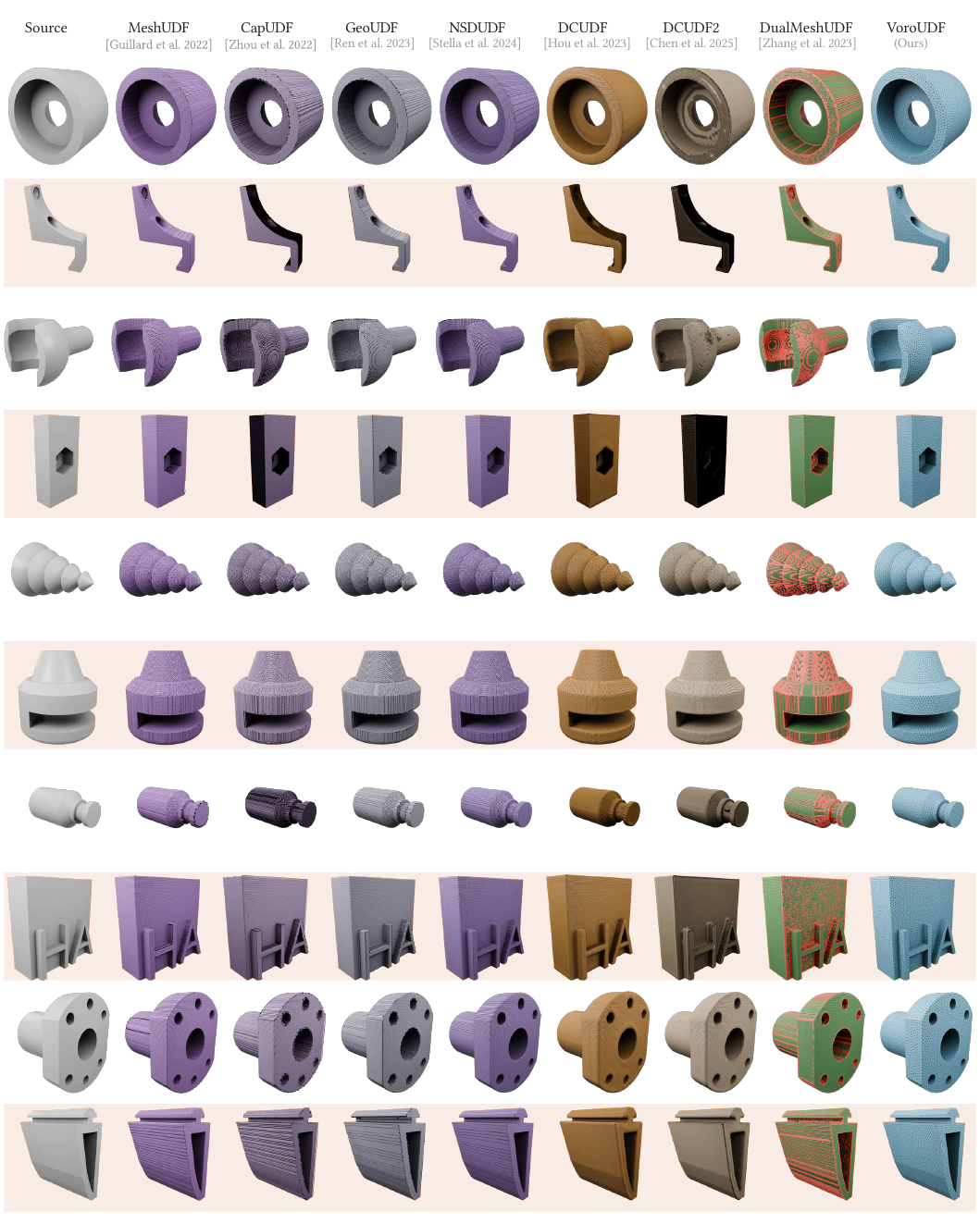}
    \vspace{-10pt}
    \caption{\rewritten{Qualitative comparison on 97 ABC models (part 1), evaluated against seven state-of-the-art methods at higher resolution ($2^7$). Non-manifold edges are highlighted in red.}}
    \label{fig:more_abc_d7_NM1_1}
\end{figure*}

\begin{figure*}
    \centering
    \includegraphics[width=0.96\linewidth]{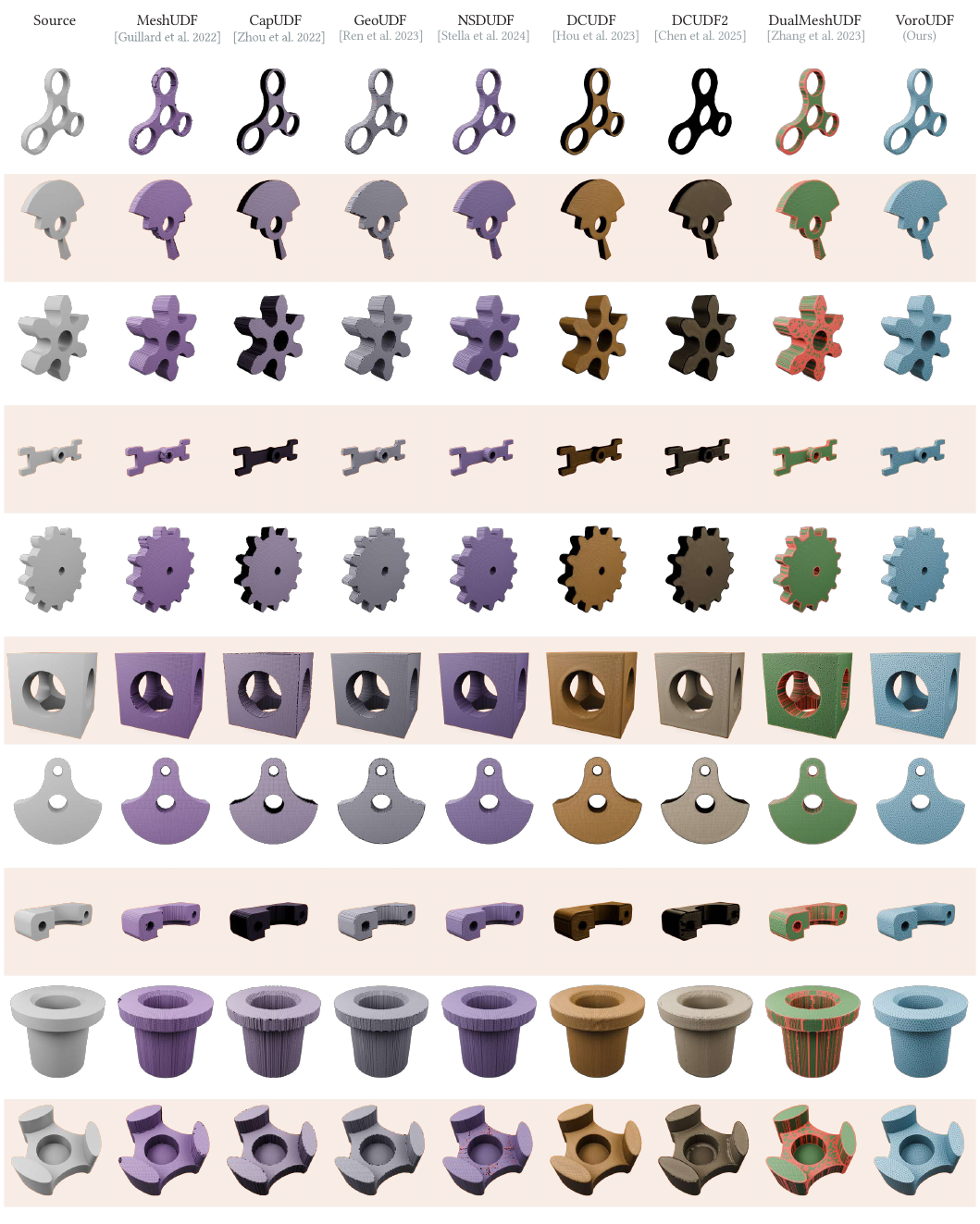}
    \vspace{-10pt}
    \caption{\rewritten{Qualitative comparison on 97 ABC models (part 2), evaluated against seven state-of-the-art methods at higher resolution ($2^7$). Non-manifold edges are highlighted in red.}}
    \label{fig:more_abc_d7_NM1_2}
\end{figure*}

\begin{figure*}
    \centering
    \includegraphics[width=0.98\linewidth]{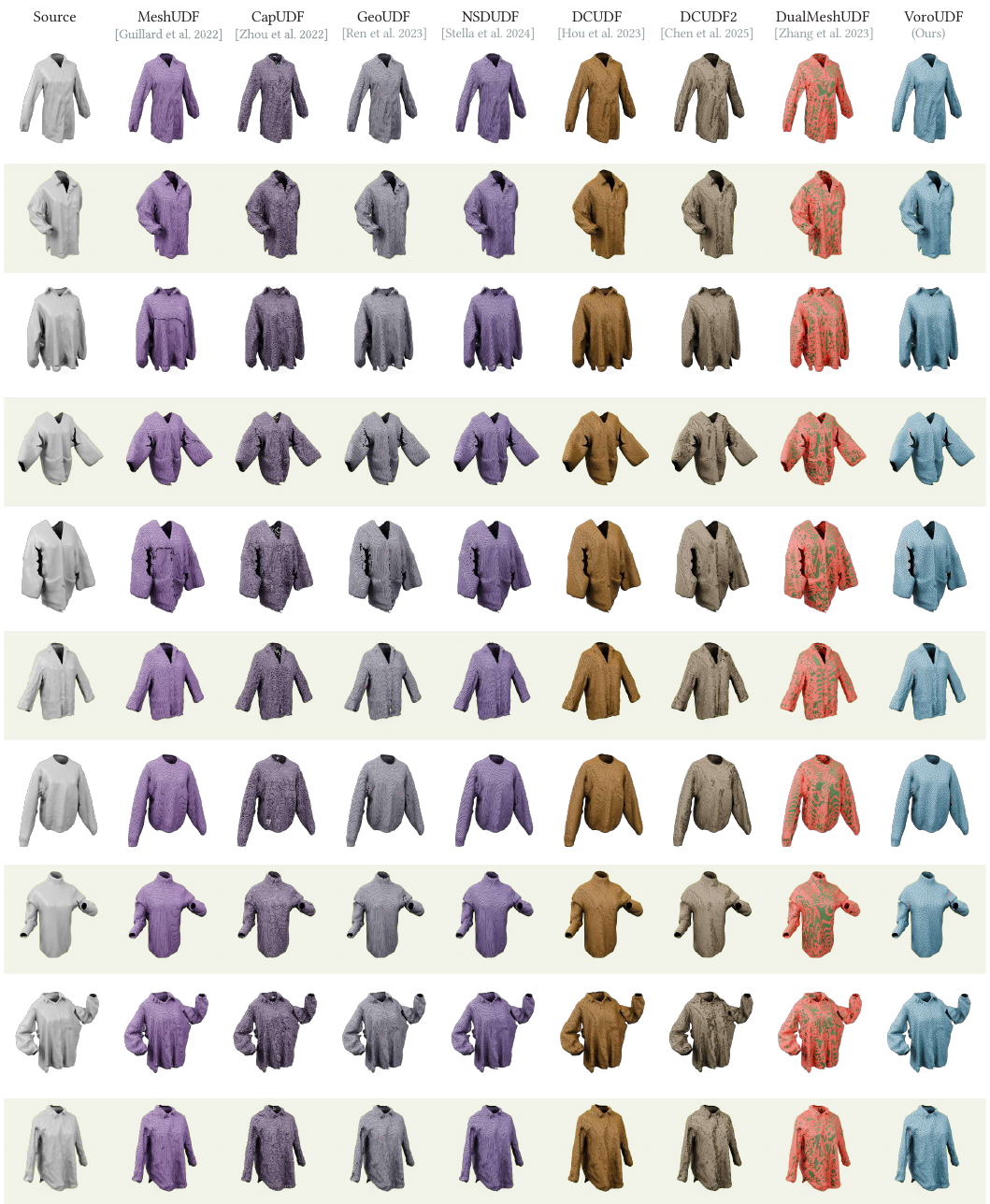}
    \vspace{-10pt}
    \caption{\rewritten{Qualitative comparison on 101 DeepFashion3D models (part 1), evaluated against seven state-of-the-art methods at higher resolution ($2^7$). Non-manifold edges are highlighted in red.}}
    \label{fig:more_deepFashion3D_d7_NM1_1}
\end{figure*}

\begin{figure*}
    \centering
    \includegraphics[width=0.98\linewidth]{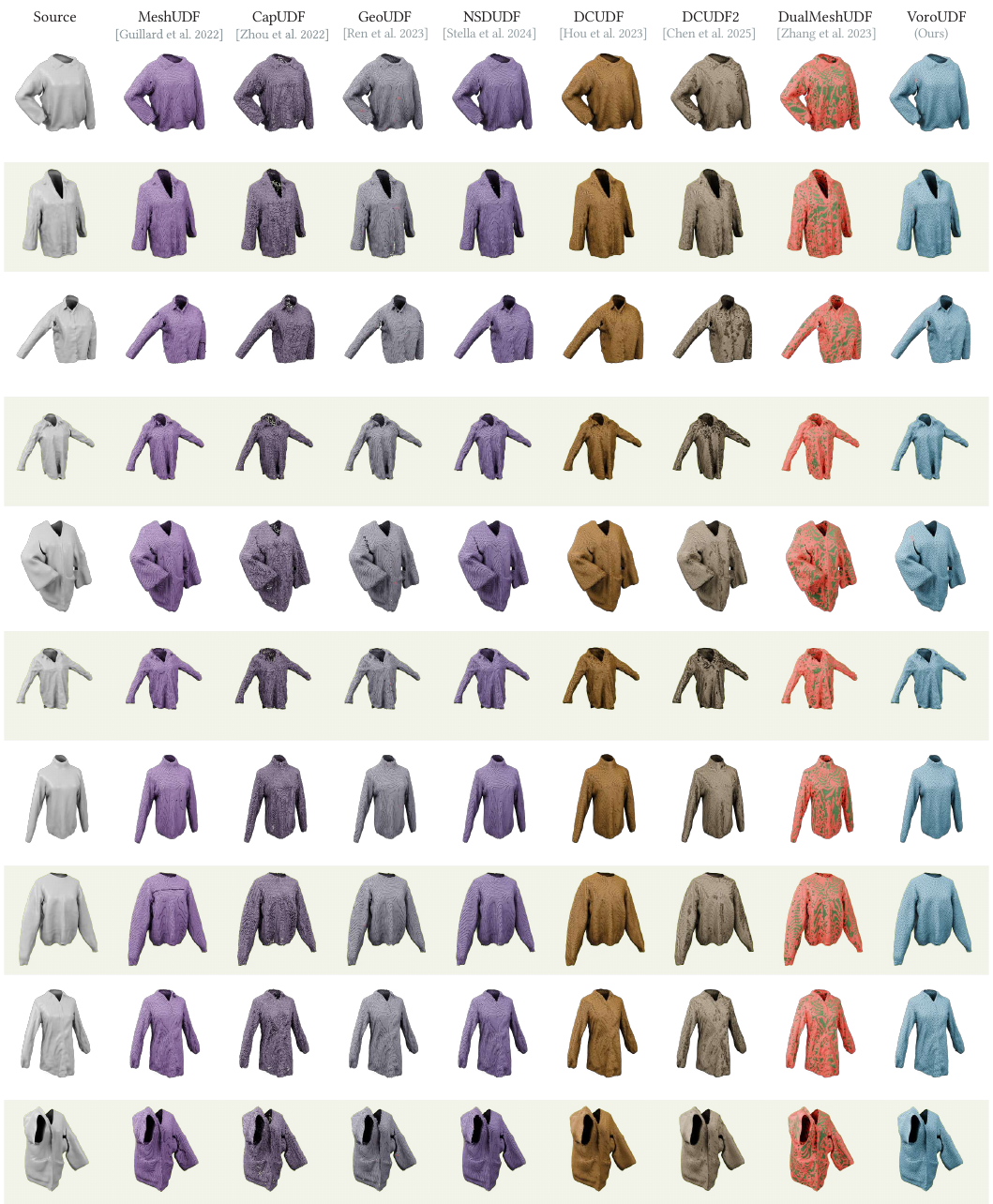}
    \vspace{-10pt}
    \caption{\rewritten{Qualitative comparison on 101 DeepFashion3D models (part 2), evaluated against seven state-of-the-art methods at higher resolution ($2^7$). Non-manifold edges are highlighted in red.}}
    \label{fig:more_deepFashion3D_d7_NM1_2}
\end{figure*}

%% file: tables/tab_exp_non_manifold.tex
\begin{table*}[t]
\caption{Quantitative comparison on 10 synthetic models with
non-manifold edges against seven state-of-the-art methods, each evaluated
at resolutions $2^5$ and $2^7$; our Voronoi-based
method has no grid resolution, so its single run is the last row. \rewritten{The
\best{best} scores are shown in bold and underlined, and the
\second{second best} scores in bold.} For prior methods
produce outputs without non-manifold structure, we mark $\nmcd$ as Inf.}
\vspace{-10pt}
\begin{center}
\resizebox{\linewidth}{!}{%
\small\setlength{\tabcolsep}{9.7pt}%
\rowcolors{2}{gray!10}{white}
\begin{tabular}{c|c||c|c|c|c|c|c|c|c}
\rowcolor{header2color}
\toprule
  Method &Res. &\#v & $\cd$ x$10^{-3}$~$\downarrow$ & $\hd$ x$10^{-3}$~$\downarrow$ & $\ecd$ x$10^{-2}$~$\downarrow$ &$\triangleQ$~$\uparrow$ & $\td$ ~$\downarrow$ & $\nmcd$ x$10^{-3}$~$\downarrow$ & Time (s) \\
\midrule
MeshUDF &$2^5$
    &7.5k &65.61 &33.08 &10.26 &0.65 &157.2 &849.37 &23.5 \\
CapUDF &$2^5$
    &21.1k &73.22 &126.87 &7.18 &0.57 &5215.9 &Inf &34.8 \\
GeoUDF &$2^5$
    &4.5k &19.77 &36.17 &25.00 &0.65 &142.1 &1448.48 &35.0 \\
NSDUDF &$2^5$
    &3.6k &13.93 &22.08 &96.69 &0.64 &84.7 &117.73 &6.4 \\
DCUDF &$2^5$
    &8.3k &90.06 &32.27 &89.70 &0.72 &\second{45.1} &Inf &954.2 \\
DCUDF2 &$2^5$
    &9.1k &69.13 &28.13 &95.95 &0.72 &\second{45.1} &Inf &1763.8 \\
DualMeshUDF &$2^5$
    &5.1k &6.33 &11.35 &0.57 &0.52 &744.3 &233.39 &15.8 \\
    \midrule
    \midrule
MeshUDF &$2^7$
    &121.1k &15.13 &7.49 &15.97 &0.66 &479.7 &1339.27 &160.3 \\
CapUDF &$2^7$
    &398k &7.85 &9.31 &6.69 &0.59 &100k &Inf &652.4 \\
GeoUDF &$2^7$
    &71.2k &4.89 &8.55 &15.34 &0.67 &521.9 &735.44 &437.5 \\
NSDUDF &$2^7$
    &60.6k &4.61 &4.07 &50.95 &0.65 &257.3 &541.67 &27.5 \\
DCUDF &$2^7$
    &159.4k &19.20 &6.78 &104.24 &0.72 &45.5 &Inf &2372.5 \\
DCUDF2 &$2^7$
    &175.6k &12.16 &5.80 &126.55 &0.72 &45.5 &Inf &3956.2 \\
DualMeshUDF &$2^7$
    &84.3k &\second{4.41} &\best{1.35} &\second{0.30} &0.61 &22k &264.18 &299.8 \\
MIND &$2^7$
    &62.2k &22.95 &77.04 &40.52 &\second{0.76} &67.6 &\second{39.83} &228.5 \\
    \midrule
    \midrule
Ours &/
    &8k &\best{4.35} &\second{3.63} &\best{0.28} &\best{0.84} &\best{43.7} &\best{0.22} &159.0 \\
\bottomrule
\end{tabular}}
\end{center}
\label{tab:exp_tab_non_manifold}
\end{table*}

%% file: tables/tab_exp_ABC.tex
\begin{table*}[t]
\caption{Quantitative comparison on 100 CAD models~\cite{xu2024cwf} against seven
state-of-the-art methods, each evaluated at resolutions $2^5$
and $2^7$; our Voronoi-based method has no grid resolution, so
its single run is the last row. \rewritten{
\best{Best} scores are shown in bold and underlined,
\second{second best} scores in bold.}}
\vspace{-10pt}
\begin{center}
\resizebox{\linewidth}{!}{%
\small\setlength{\tabcolsep}{13.9pt}%
\rowcolors{2}{gray!10}{white}
\begin{tabular}{c|c||c|c|c|c|c|c|c}
\rowcolor{header2color}
\toprule
  Method &Res. &\#v & $\cd$ x$10^{-3}$~$\downarrow$ & $\hd$ x$10^{-3}$~$\downarrow$ & $\ecd$ x$10^{-2}$~$\downarrow$ &$\triangleQ$~$\uparrow$ & $\td$ ~$\downarrow$ & Time (s) \\
\midrule
MeshUDF &$2^5$
    &4.2k &61.89 &28.40 &25.13 &0.66 &50.2 &10.2 \\
CapUDF &$2^5$
    &6.7k &60.38 &94.71 &16.68 &0.52 &193.2 &11.4 \\
GeoUDF &$2^5$
    &2.6k &19.29 &31.58 &41.53 &0.67 &62.4 &11.3 \\
NSDUDF &$2^5$
    &2.2k &13.96 &15.87 &100.35 &0.65 &17.1 &5.8 \\
DCUDF &$2^5$
    &6.6k &91.33 &31.87 &110.37 &0.65 &5.2 &474.0 \\
DCUDF2 &$2^5$
    &8.5k &72.51 &27.78 &112.16 &0.62 &5.3 &823.1 \\
DualMeshUDF &$2^5$
    &3k &6.88 &10.61 &2.31 &0.53 &691.2 &5.8 \\
    \midrule
    \midrule
MeshUDF &$2^7$
    &69.7k &14.06 &6.26 &29.61 &0.67 &242.4 &72.0 \\
CapUDF &$2^7$
    &111.3k &7.01 &6.16 &11.16 &0.53 &816.6 &653.0 \\
GeoUDF &$2^7$
    &41.8k &3.95 &6.59 &27.61 &0.68 &300.2 &18.0 \\
NSDUDF &$2^7$
    &35.6k &3.72 &2.79 &54.96 &0.66 &97.5 &11.9 \\
DCUDF &$2^7$
    &128.8k &19.37 &5.89 &130.18 &0.64 &4.8 &1087.9 \\
DCUDF2 &$2^7$
    &179.9k &13.53 &5.00 &103.06 &0.61 &\second{4.7} &1786.7 \\
DualMeshUDF &$2^7$
    &48.3k &\second{3.37} &\best{0.98} &\second{0.85} &0.61 &10k &64.6 \\
MIND &$2^7$
    &35.6k &16.38 &22.47 &123.52 &\second{0.73} &12.2 &141.3 \\
    \midrule
    \midrule
Ours &/
    &6.4k &\best{3.32} &\second{2.40} &\best{0.11} &\best{0.85} &\best{3.8} &90.7 \\
\bottomrule
\end{tabular}}
\end{center}
\label{tab:exp_tab_ABC}
\end{table*}

%% file: tables/tab_exp_deepFashion3D.tex
\begin{table*}[t]
\caption{Quantitative comparison on top 100 garment models from
deepFashion3D~\cite{zhu2020deepfashion} against eight state-of-the-art
methods, each evaluated at
resolutions $2^5$ and $2^7$; our Voronoi-based
method has no grid resolution, so its single run is the last row. \rewritten{The
\best{best} scores are shown in bold and underlined, and the
\second{second best} scores in bold.}}
\vspace{-10pt}
\begin{center}
\resizebox{\linewidth}{!}{%
\small\setlength{\tabcolsep}{17.9pt}%
\rowcolors{2}{gray!10}{white}
\begin{tabular}{c|c||c|c|c|c|c|c}
\rowcolor{header2color}
\toprule
  Method &Res. &\#v & $\cd$ x$10^{-3}$~$\downarrow$ & $\hd$ x$10^{-3}$~$\downarrow$ &$\triangleQ$~$\uparrow$ & $\td$ ~$\downarrow$ & Time (s) \\
\midrule
MeshUDF &$2^5$
    &4k &66.18 &31.36 &0.61 &77.5 &26.5 \\
CapUDF &$2^5$
    &18.2k &29.07 &30.90 &0.54 &4609.1 &26.2 \\
GeoUDF &$2^5$
    &2.4k &15.46 &31.78 &0.61 &77.7 &26.9 \\
DCUDF &$2^5$
    &5.2k &91.67 &35.49 &0.72 &\second{3.3} &1057.1 \\
DCUDF2 &$2^5$
    &6k &74.49 &32.53 &0.71 &\second{3.3} &2014.7 \\
DualMeshUDF &$2^5$
    &3.1k &8.41 &27.01 &0.52 &497.9 &12.7 \\
    \midrule
    \midrule
MeshUDF &$2^7$
    &66.9k &15.91 &8.20 &0.61 &727.6 &188.2 \\
CapUDF &$2^7$
    &305.7k &6.54 &9.05 &0.53 &78k &877.6 \\
GeoUDF &$2^7$
    &37.4k &3.54 &10.07 &0.62 &238.9 &601.6 \\
DCUDF &$2^7$
    &100.6k &19.21 &7.90 &0.73 &7.7 &2288.8 \\
DCUDF2 &$2^7$
    &112.3k &12.58 &\second{7.14} &0.71 &7.5 &3713.9 \\
DualMeshUDF &$2^7$
    &51.1k &\best{3.27} &7.62 &0.58 &8130.1 &67.0 \\
MIND &$2^7$
    &35k &5.78 &26.85 &\second{0.77} &23.6 &270.0 \\
    \midrule
    \midrule
Ours &/
    &7.7k &\second{3.49} &\best{5.47} &\best{0.83} &\best{2.1} &122.8 \\
\bottomrule
\end{tabular}}
\end{center}
\label{tab:exp_tab_deepFashion3D}
\end{table*}